\documentstyle[12pt,fvtijmp2]{article}
\begin{document}

\def\eq#1{(\ref{#1})}
\def\Eq#1{Eq.(\ref{#1})}
\def\Eqs#1-#2{Eqs.\eq{#1}--\eq{#2}}
\def\eqs#1-#2{\eq{#1}--\eq{#2}}

\def\SECTION{\section}
\def\SUBSECTION{\subsection}

\def\asop{$As$-operation}
\def\r*op{$R^{*}$-operation}
\def\rop{$R$-operation}
\def\asy#1#2{\ \displaystyle\mathop\simeq_{
\mathstrut #1\rightarrow#2}\ } 

\def\hphi{\hat\varphi}
\def\tphi{\tilde\varphi}
\def\tj{\tilde{j}}
\def\bold#1{{\bf#1}}
\def\.{\raise0.3ex\lbox{\scriptscriptstyle\circ}}
\def\lbox#1{\hbox{\mathsurround0pt$#1$}}

\thispagestyle{empty}
\vbox to1in{\vfill}         
\hfill \hbox{FERMILAB-PUB-91/345-T}
\begin{center}

{\large
    EUCLIDEAN ASYMPTOTIC EXPANSIONS\\
[2mm]
   OF GREEN FUNCTIONS OF QUANTUM FIELDS.\\
[2mm]
    (II) {\sc Combinatorics of the asymptotic operation}\\
[6mm]}  
{G.~B.~Pivovarov} \\
[2mm]
Institute for Nuclear Research \\
of the Russian Academy of Sciences, \\
Moscow 117312, Russia. \\
[3mm]
and \\
[3mm]
{ F.~V.~Tkachov}%
\footnote{on leave from Institute for Nuclear Research of 
the Russian Academy of Sciences, Moscow 117312, Russia}
\\
[2mm]
Fermi National Accelerator Laboratory \\
P.O.Box 500, Batavia, Illinois 60510 USA\\
[15mm]

\end{center}

{
\centerline{\sc ABSTRACT}
\vskip2mm
\noindent
The results of \cite{partI} are used to obtain full asymptotic 
expansions of Feynman diagrams renormalized within the MS-scheme in 
the regimes when some of the masses and external momenta are large 
with respect to the others. The large momenta are Euclidean, and the 
expanded diagrams are regarded as distributions with respect to them. 
The small masses may be equal to zero. The \asop\ for integrals is 
defined and a simple combinatorial techniques is developed to study its 
exponentiation. The 
\asop\ is used to obtain the corresponding expansions of arbitrary 
Green functions. Such expansions generalize and improve upon the 
well-known short-distance operator-product expansions, the decoupling 
theorem etc.; e.g.\ the low-energy effective Lagrangians are obtained 
to all orders of the inverse heavy mass. The obtained expansions 
possess the property of perfect factorization of large and small 
parameters, which is essential for meaningful applications to 
phenomenology. As an auxiliary tool, the inversion of the \rop\ is 
constructed. The results are valid for arbitrary QFT models.
}

\setcounter{footnote}{0}
\newpage

\newpage\thispagestyle{myheadings}\markright{}
\section{Introduction}
\label{introduction}

In the  preceding paper \cite{partI} the problem of obtaining 
asymptotic expansions of multiloop Feynman diagrams was analyzed. From 
the point of view of phenomenology, it was stressed that such 
expansions must possess the property of ``perfect factorization" of 
large and small dimensional parameters. From the 
point of view of mathematics, it was shown that the key technical 
problem (``the Master problem") is that of obtaining asymptotic 
expansions---in powers and logarithms of the expansions parameter---of 
products of singular functions in the sense of distributions. 
An explicit solution for the Master problem was obtained in 
the form of the so-called {\em \asop\ for products of singular 
functions,} which acts on Euclidean momentum-space Feynman integrands 
and yields an expansion whose all terms are well-defined distributions 
(containing, in particular, $\delta$-functional contributions with 
coefficients depending non-analytically on the expansion parameter) so 
that the expansion allows termwise integrations with test functions.

The purpose of the present paper is to use the results of \cite{partI} 
and obtain explicit expansions of multiloop Feynman integrals in the 
form of {\em \asop\ for integrals}.%
\footnote{
The two \asop s are different representations of essentially the same 
operation, which justifies using the same name to denote them: the 
\asop\ for integrals is an integrated version of the \asop\ for 
products---cf.\ subsect.~\ref{s4}.1 below.
} 
It turns out that the combinatorial structure of the \asop\ for 
integrals is very similar to that of the \rop%
\footnote{
Note that the \asop\ for products also has close parallels with the  
\rop---but with the \rop\ in position representation.}.
This similarity allows one to derive expansions of perturbative Green 
functions in a global form with the help of the same techniques as 
used in studying the exponentiation of the \rop. 

The present paper is a revised and simplified version of our 
publications \cite{IIold'}, \cite{IIold"} and \cite{Rinv} 
which have been widely discussed 
in the literature \cite{gorishny}--\cite{MPI}.%
\footnote{
OPE in the MS-scheme was also studied in \cite{llew}.
\relax}
The revision
has not affected the results but only the  order of presentation.
The simplification is due to the fact that the  starting point of
\cite{IIold'} was the so-called EA-expansion for UV-renormalized
integrals, which is a composition of $R$-operation and the
$As$-operation in the sense of the present paper.  Correspondingly,
the \asop\ as defined in \cite{IIold'} mixed up  UV-renormalization
and the expansion algorithm. In the present version  we define the
\asop\ and study its exponentiation directly for the  unrenormalized
integrals, which results in further simplifications. 
The attractiveness of the \asop\ is due to the fact that 
it serves as a combinatorial 
organizing center of the theory and fully exhibits the structure of
the expansions, which is invaluable for applications where 
diagram-by-diagram analysis is necessary to enhance reliability of 
calculations. Its definition is also remarkably simple within 
dimensional regularization, which trivializes the study of 
exponentiation. It should be stressed, however, that the use of 
dimensional regularization is by no means essential for the expansion
problem; a separate publication is devoted to a study of this
point---cf. \cite{fvt-vvv}, \cite{gms}, and especially \cite{IV}. In
the latter work Euclidean asymptotic expansions were studied in
regularization independent manner within the framework of generalized
MS-schemes
\cite{gms}. The main result of \cite{IV} is an expansion formula whose
combinatorial structure is practically identical to that of the $As$-operation 
of the present paper. Therefore, all our combinatorial results 
concerning exponentiation of the $As$-operation are immediately 
applicable to \cite{IV}.

The plan of the present paper is as follows. It consists of five 
sections, each section is subdivided into subsections and contains a 
prologue where further information can be found. In the first three 
sections various aspects of combinatorial structure of the \rop\ in 
the MS-scheme are studied. In sect.~\ref{s1} the exponentiation formula 
for the \rop\ is obtained using the functional techniques. The result 
is, of course, well-known. However, since we wish to use similar 
arguments in studying the \asop, we feel it is useful to present an 
explicit derivation in such a way as to make it immediately applicable 
to the case of the \asop\ studied in sects.~4 and 5. In sect.~\ref{s2} 
the inversion of the \rop\ is derived, and in sect.~\ref{s3} the 
renormalization of Green functions with multiple local operator 
insertions is considered.

In sects.~4 and 5 we study the combinatorial structure of asymptotic 
expansions of Feynman diagrams. In sect.~\ref{s4} we consider the 
special case of heavy mass expansions. The \asop\ for integrals is 
introduced and its exponentiation is proved. Using the techniques 
developed in sects.~1--3, the effective low-energy Lagrangian is 
presented in an explicitly convergent form to all orders in the 
inverse heavy mass. (It should be stressed that in this paper we only 
deal with purely combinatorial aspects of the theory. A fuller 
justification of a technical assumption (see subsect.~\ref{ss4.8} 
below) is presented in \cite{IV} (for an informal discussion see \cite{III}.) 
In sect.~\ref{s5} the results of sect.~\ref{s4} are extended to comprise 
the case of large external momenta, so that the familiar 
short-distance operator product expansion is reproduced and its 
generalizations are obtained.

Note that our final formulae immediately provide the very useful 
explicit expressions of OPE coefficient functions first announced in 
\cite{alg:83} (explicit examples of calculations going as far as 
3-loop approximation can be found in \cite{ope3loops}; for further 
references see also \cite{paradigm}).

The notations are on the whole consistent with those of \cite{partI}.

\newpage\thispagestyle{myheadings}\markright{}
\centerline{{\large \bf COMBINATORICS OF THE $R$-OPERATION}}
\centerline{\hbox{}}

\centerline{{\large \bf IN THE MS-SCHEME}}

\section{$R$-operation in the MS-scheme}
\label{s1}

In this section we study the combinatorial structure of the \rop\
applied to the perturbation series as a whole. The fact that the \rop\
is equivalent to adding counterterms to the Lagrangian is of course
well-known. But we are interested in exact formulae expressing this
fact that would be convenient for practical calculations and valid for
the MS-scheme of UV renormalization. Therefore we reproduce a simple
proof of the exponentiation of the \rop\ in the MS-scheme---see
\eq{(1.26)} below---as given in \cite{IIold'}. Our second aim is to
present the derivation of the exponentiation formula \eq{(1.26)} in a
form sufficiently general to make it applicable to the case of the
\asop\ studied in sects.~4 and 5. To make the paper more accessible
to practitioners of applied QFT, our presentation is rather detailed.

We start in subsect.~\ref{ss1.1} with a description of the \rop\ in the 
MS-scheme. Our definitions are equivalent to the standard ones 
\cite{collins}, and the differences are mainly notational. In 
subsect.~\ref{ss1.2} a convenient order of enumeration of UV-subgraphs 
is fixed in preparation to studying the global structure of the \rop. 
Subsect.~\ref{ss1.3} summarizes some results of the functional 
techniques that will be needed (for more details see e.g.\ 
\cite{funcMethods}). In subsect.~\ref{ss1.4} a formal expression of the 
statement that the \rop\ is equivalent to adding counterterms to the 
Lagrangian, the exponentiation formula \eq{(1.26)}, is derived; our 
reasoning is an extension of the arguments of \cite{zav}---for 
comments and comparison see subsect.~\ref{ss1.5}). In 
subsect.~\ref{ss1.6} the results are extended to Green functions of 
fields. Complications due to composite operator insertions will be 
discussed in sect.~\ref{s3}.

\subsection{Basic definition of the \rop}
\label{ss1.1}

The effect of the \rop\ in the MS-scheme on a single Feynman integral 
$G$ can be expressed by the formula
\begin{equation}\label{(1.1)} 
     \bold{R} \. G 
     = \sum_{\{ g_\alpha \}} 
          \left( \prod_{\alpha} \Delta _{\rm UV}\. g_{\alpha} \right)
          \times G / \left( \prod_{\alpha} g_{\alpha} \right).
\end{equation}
Let us explain it.

One associates a Feynman graph, that will be referred to as ``the graph 
$G$", with the integral $G,$ and vice versa. The correspondence 
between graphs and integrals is established by Feynman rules.

Let us fix how $\mu,$ the 't Hooft unit of mass, appears in the 
integrals. In momentum representation, having taken into account 
momentum conservation in all the vertices of the graph, one can obtain 
the integrand of the corresponding unrenormalized Feynman integral. 
Then for each internal momentum integration one replaces the measure 
as
\begin{equation}\label{(1.2)} 
     {d^4p \over (2\pi)^4} \to \mu^{2\epsilon} 
     {d^Dp \over (2\pi)^D},     
     \quad\quad 2\epsilon = 4-D,
\end{equation}
and assumes that integrations (as well as the Dirac algebra etc.) are to 
be done within dimensional regularization. The role of $\mu $ is to 
preserve the dimensionality of the integral as a whole under 
regularization.

An {\em UV-subgraph} $g$ of $G$ consists of some vertices of the graph 
$G$ and some of its lines attached by both ends to the vertices of 
$g.$ If $g$ consists of one vertex and no lines then we call it {\em 
elementary vertex.} If $g$ has at least one loop and is connected and 
one-particle-irreducible then we call $g$ a {\em proper UV-subgraph.} 
(Note that if $G$ is one-particle-irreducible then one of its proper 
UV-subgraphs coincides with $G.$)

An {\em UV-partition} of $G$ is a set of its UV-subgraphs 
$\{g_{\alpha}\}$ such that each vertex of $G$ belongs to one and only 
one $g_{\alpha}.$ Note that one of the UV-partitions consists of 
elementary vertices only, and another one consists of only one 
subgraph, namely, $G$ as a whole. Summation in \eq{(1.1)} runs over 
all UV-partitions of $G$.

Now we have to define the operation $\Delta _{\rm UV}\. g$ where $g$ 
is an UV-subgraph of $G$. By definition, $\Delta _{\rm UV}\. g = g$ 
if $g$ is an elementary vertex, and $\Delta _{\rm UV}\. g = 0$ if $g$ 
is not an UV-subgraph. If $g$ is a proper UV-subgraph, then the action 
of $\Delta _{\rm UV}$ on $g$ can be described as follows. The graph 
for $(\Delta _{\rm UV}\. g)\times (G/g)$ can be obtained from the graph 
for $G$ by shrinking $g$ in $G$ to a point. With the new vertex thus 
obtained one associates a factor---the UV-counterterm---that is 
obtained from the integral corresponding to $g$ via a special 
algorithm which need not be specified here (see, however, below). 
Using the Feynman rules thus extended one finally builds up the 
integral for $(\Delta _{\rm UV}\. g)\times (G/g).$ Note that the above 
rule for $\mu$  is still operative, i.e.\ the number of the factors 
$\mu ^{2\epsilon}$ is equal to the number of loops of $G$ with $g$ 
shrunk to a point.

For the sake of completeness let us describe the recipe for evaluating 
the UV-counterterm for a given proper UV-subgraph $g.$ If one rewrites 
the \rop\ as
\begin{equation}\label{(1.3)} 
     \bold{R} \. g \equiv \bold{R}'\. g + \Delta _{\rm UV}\. g ,
\end{equation}
then $\Delta _{\rm UV}\. g$ is precisely the UV-counterterm to be 
evaluated while $\bold{R} '$ involves only counterterms for subgraphs 
of $g$ which are assumed to be known already. Since we are working 
within dimensional regularization, all our integrals depend 
parametrically on $D,$ the complex-valued dimension of space-time. Let 
$\bold{K}$ be the operation that picks out the pole part at $D=4$ of 
any function on which it acts. $\bold{R} \. g$ should be finite at $D 
=4,$ therefore $\bold{K}\. \bold{R} \. g = 0.$ On the other hand, 
the UV-counterterms in the MS-scheme are pure poles. So, applying 
$\bold{K}$ to \eq{(1.3)}, we get:
\begin{equation}\label{(1.4)} 
     \Delta_{\rm UV} \. g = - \bold{K} \. \bold{R}' \. g.
\end{equation}
Eqs.~\eq{(1.1)}, \eq{(1.3)} and \eq{(1.4)} provide a convenient 
description of the \rop\ in the MS-scheme. We would like to stress 
that all the manipulations are to be done in momentum representation, 
and that the rule for $\mu$ remains operative. Also recall that 
\eq{(1.4)} is a polynomial of masses and external momenta of $g$ and 
is independent of $\mu.$

The most important points for us here are:

($i$) $\Delta_{\rm UV} \. g,$ whatever the subgraph $g$ is, depends 
only on $g$ but not on $G$ as a whole.

($ii$) Summation in \eq{(1.1)} runs over all partitions of $G.$

($iii$) An UV-subgraph is defined as a subset of vertices and some 
(not necessarily all) of the lines connecting the vertices of this 
subset.

On the other hand, the fact that $\Delta_{\rm UV}$ is non-trivial only 
on {\em proper} UV-subgraphs will be completely irrelevant.

\subsection{Enumerating UV-subgraphs}
\label{ss1.2}

Let us transform \eq{(1.1)} to a more convenient form. Indeed, there are
UV-subgraphs that have the same set of vertices but differ only 
with respect of the number and/or arrangement of lines 
connecting them (e.g.\ when a line of a subgraph can be removed 
without changing its status as 1PI)---but not vice versa: if the sets of 
lines of two subgraphs are identical, than their sets of vertices are 
identical, too.
This allows one to perfrom enumeration of UV-subgraphs in \eq{(1.1)} 
in two stages:
first, one enumerates sets of vertices, then, sets of lines connecting 
those vertices.

Indeed, let $v$ be a subset of vertices of $G$ and
\begin{equation}\label{(1.5)} 
     \Lambda _{v} \. G \equiv \sum_{g\sim v} 
     ( \Delta _{\rm UV} \. g ) \times (G/g),
\end{equation}
where summation runs over all UV-subgraphs $g$ with the same set of 
vertices, $v$. Denoting the set of all vertices of $G$ as $V$, we can 
rewrite \eq{(1.1)} as:
\begin{equation}\label{(1.6)}
     \bold{R} \. G 
     = \mathop{\sum_{v=\cup v_\alpha}
              }_{v_\alpha\cap v_\beta=\emptyset,
                           {\ \rm for\ }
                           \alpha\not=\beta 
                }
                     \left( \prod_{\alpha} \Lambda_{v_{\alpha}} 
                     \right) 
                     \. G.
\end{equation}

\subsection{$T$-products and functional techniques}
\label{ss1.3}

Let $\hphi(x)$ be a free field operator; $\varphi (x)$ will denote the 
corresponding classical field. Let $F(\varphi)$ be a functional of the 
form
\begin{equation}\label{(1.7)} 
     F(\varphi ) 
     = \sum_n \int F_{n}(x_1\ldots x_{n}) 
           \varphi (x_1)\ldots\varphi (x_{n})
            dx_1\ldots dx_{n}.
\end{equation}
Replacing the product of $\varphi $'s in \eq{(1.7)} by the $T$-product 
of $\hphi$'s one obtains an operator which we denote as ${\rm 
T}F(\hphi).$ The Wick theorem allows one to reexpand ${\rm T}F(\hphi)$ 
in terms of the Wick normal products of $\hphi:$
\begin{equation}\label{(1.8)} 
     {\rm T}F(\hphi) = \sum_n \int F^{\rm N}_{n}(x_1\ldots x_{n}) 
          \,{\rm N}\left[\hphi(x_1)\ldots\hphi(x_{n})\right] 
          dx_1\ldots dx_{n},
\end{equation}
where $F^{\rm N}_{n}$ are related to $F_{n}$ in a certain way (see 
below).

The generating functional for the coefficient functions is defined as:
\begin{equation}\label{(1.9)}
     F^{\rm N}(\varphi ) = \sum_n \int 
       F^{\rm N}_{n}(x_1\ldots x_{n}) 
       \varphi (x_1)\ldots\varphi (x_{n})
        dx_1\ldots dx_{n}.
\end{equation}
The relation between $F(\varphi)$ and $F^{\rm N}(\varphi)$ is 
expressed, by definition, by
\begin{equation}\label{(1.10)} 
     {\rm T}F(\hphi) = {\rm N}[F^{\rm N}(\varphi)]_{\varphi=\hphi}. 
\end{equation}
The well-known formal expression of the Wick theorem due to Khori (see 
\cite{funcMethods} and refs.\ therein) reads:
\begin{equation}\label{(1.11)}
     F^{\rm N}(\varphi ) = e^l\, F(\varphi ),
\end{equation}
where
\begin{equation}\label{(1.12)}  
     l \equiv {1\over2}\int\,dx\,dy\,
     {\delta\over\varphi(x)} \Delta^c(x,y) {\delta\over\varphi(y)}
     \equiv {1\over2}\delta\Delta^c\delta,
\end{equation}
and
\begin{equation}\label{(1.13)}
     \Delta^c(x,y) = <{\rm T}\hphi(x)\hphi(y)>_0.
\end{equation}
(Extension of all our formulae to the most general case of fermionic 
and complex fields is straightforward, therefore our analysis will be 
quite general.)

Our purpose is to study Green functions, but first it is convenient to 
consider the (off-shell) $S$-matrix. The unrenormalized $S$-matrix is 
expressed as:
\begin{equation}\label{(1.14)}
     {\rm S}(\hphi ) = {\rm T}\exp[L(\hphi)],
\end{equation}
where $L(\hphi )$ denotes the interaction Lagrangian integrated over 
space-time and multiplied by $i$ (the normal ordering of fields in 
$L(\hphi )$ is {\em not} assumed).

The generating functional $S^{\rm N}$ is defined as (cf.\ \eq{(1.10)}):
\begin{equation}\label{(1.15)}
     {\rm T}\exp[L(\hphi)] 
     = {\rm N}[S^{\rm N}(\varphi)]_{\varphi=\hphi} 
\end{equation}
and from \eq{(1.11)}:
\begin{equation}\label{(1.16)}
     S^{\rm N} = e^l\exp[L(\varphi)].
\end{equation}
Again we have to set the rules for $\mu$, in order to establish 
connection between our earlier definitions of Feynman integrals and 
the expressions generated using the functional technique. We assume 
that:

($i$) all momentum and space-time integrations, $\gamma$-matrices etc. 
are $D$-dimensional;

($ii$) the dimensions of the fields $\hphi $ and the coupling 
constants are always the canonical 4-dimensional ones;

($iii$) each propagator bears an extra factor $\mu ^{2\epsilon};$ note 
that \eq{(1.12)} is dimensionless since we assume that $\delta /\delta 
\varphi (x)\times \varphi (y)=\delta (x-y)$ with $D$-dimensional 
$\delta $-function;

($iv$) the interaction Lagrangian $L(\varphi )$ in \eq{(1.14)} 
contains the factor $\mu^{-2\epsilon}$ and is therefore dimensionless; 
note that condition ($iii$) is equivalent to ($iv$) extended to the 
quadratic part of the full Lagrangian ( cf.\ \eq{(1.36)} below).

Given these rules, a connected Feynman integral will have the factor 
$\mu^{2\epsilon(l-1)}$ where $l$ is the number of its internal 
momentum integrations (loops)---which differs by $\mu ^{-2\epsilon}$ 
from what was postulated in subsect.~\ref{ss1.1}. Such extra $\mu 
^{-2\epsilon}$ to each connected component can be easily taken into 
account in what follows.

\subsection{Exponentiation of the \rop}
\label{ss1.4}
\label{ss1.5}

The renormalized $S$-matrix is obtained by applying the \rop\ to each 
Feynman integral contributing to the coefficient functions of $S^{\rm 
N}.$ Thus, the starting point of our analysis is the following 
expression:
\begin{equation}\label{(1.18)}
     {\rm S}_{\rm R} (\varphi ) 
     = \bold{R} \. [{\rm T}\,\exp(L(\varphi ))] 
     = \sum_{N=0}^\infty{1\over N!} \bold{R} \. {\rm T} L^N(\varphi ),
\end{equation}
where we have introduced the convenient notation 
\begin{equation}\label{(1.19)}
     {\rm T} \equiv e^l,
\end{equation}
which will be used systematically in the context of the functional 
techniques.

Evaluating functional derivatives, one obtains a sum of terms, each 
one having a graphic representation. In the resulting expression one 
only has to replace products of the classical fields $\varphi $ by the 
normal ($N$-) products of the free fields $\hphi ,$ in order to obtain 
the operator of the $S$-matrix. (Note that since we wish to use the 
MS-scheme for UV renormalization, the $S$-matrix will not satisfy 
correct normalization conditions automatically, but this is of no 
importance because our study of the $S$-matrix is only an intermediate 
step in the study of Green functions.)

Note that in \eq{(1.18)} each $L(\varphi )$ corresponds to a vertex 
while each $l$ (see \eq{(1.12)}) generates one line of a Feynman 
graph. This allows one to conveniently perform all the enumerations of 
UV-partitions inherent in the definition of the \rop\ as given by 
eqs.~\eq{(1.1)}--\eq{(1.4)}.

First one has to enumerate all possible partitions of $N$ vertices 
into groups of vertices of various sizes (cf.\ \eq{(1.6)}). $N$ 
vertices can be split into $n_1$ groups of 1 vertex, $n_2$ groups of 2 
vertices $\ldots $ $n_N$ groups of $N$ vertices in
\begin{equation}\label{(1.20)}
     {N\atopwithdelims\{\}n_1\ldots n_N} 
     = N! \left[ \prod_{k=1}^N n_k! (k!)^{n_k} \right]^{-1},
     \quad \sum_k k \times n_k = N,
\end{equation}
ways. Therefore, applying \eq{(1.6)} to \eq{(1.18)} we get:
\begin{equation}\label{(1.21)}
     {\rm T} \sum_{N=0}^\infty {1\over N!}
     \sum_{ n_1\ldots n_N \atop \sum k n_k=N }
     {N\atopwithdelims\{\}n_1\ldots n_N} 
     \left( \Lambda \. L^1 \right)^{n_1} \ldots 
     \left( \Lambda \. L^N \right)^{n_N}.
\end{equation}
(Note the abuse of notation: $\Lambda$ acts on subgraphs consisting of
vertices {\em and\/} lines---but the latter will only be generated by
T. However, we use such a representation only for the purposes of
combinatorial enumeration so that no problems should arise.) Now, for
each group of vertices we have to expand $\Lambda $ according to
\eq{(1.5)}. We have to separate the lines connecting vertices inside
each group $\Lambda \. L^k$ from the lines connecting vertices of
different groups, and among the former to enumerate all subsets of
lines that will form UV-subgraphs on which $\Delta _{\rm UV}$ acts.
This can be done by ascribing different labels to the $n_1+\ldots+n_N$
groups of vertices and marking the field $\varphi $ in each group by
this label. Then we have:
$$
     {\rm T}\left[\prod_\alpha\Lambda\.L^{k_\alpha}(\varphi)\right] 
     = \exp({1\over2}\delta\Delta^c\delta)
     \left[ \prod_\alpha\Lambda\.L^{k_\alpha}(\varphi_{\alpha})
     \right]_{\varphi_\alpha=\varphi}
$$
$$
     = \left[ \exp
          \left( \sum_{\alpha<\beta}
                     \delta_{\alpha}\Delta^c\delta_\beta
               + \sum_\alpha {1\over2} 
                     \delta_{\alpha}\Delta^c \delta_{\alpha} 
          \right) 
          \prod_\alpha \Lambda \. L^{k_\alpha}(\varphi _{\alpha})
       \right] _{\varphi_\alpha=\varphi}
$$
\begin{equation}\label{(1.22)} 
     = \left[ \exp
          \left( \sum_{\alpha<\beta}
                     \delta_{\alpha}\Delta^c\delta_\beta
          \right) 
          \prod_\alpha \Lambda \. 
               \left( \exp({1\over2}
                      \delta_{\alpha}\Delta^c \delta_{\alpha} 
                      L^{k_\alpha}(\varphi_{\alpha})
               \right)
       \right] _{\varphi_\alpha=\varphi}
\end{equation}
(here $\delta _{\alpha}=\delta /\delta \varphi _{\alpha}$ while the
index $k_\alpha$ takes the value 1 $n_1$ times, \dots, the value $N$,
$n_N$ times). In \eq{(1.22)}, each operator $l_{\alpha}=\delta
_{\alpha}\Delta^c\delta _{\alpha}$ generates one line attached to
vertices of the $\alpha$-th group. There are ${M\choose M'} = M!
(M'!(M-M')!)^{-1}$ ways to choose $M'$ lines from a set of $M$ lines.
Therefore,
$$
     \Lambda\.(e^l\,F(\varphi)) 
     = \sum_{M=0}^\infty {1\over M!} \Lambda \. (l^M F(\varphi ))
$$
\begin{equation}\label{(1.23)}
     = \sum_{M=0}^\infty {1\over M!}
       \sum_{M'=0}^M {M\choose M'} l^{M-M'} 
       \Delta^f_{\rm UV} \. (l^{M'}\,F(\varphi )) 
     = e^l\,\Delta^f_{\rm UV} \. (e^l\,F(\varphi )), 
\end{equation}
where
\begin{equation}\label{(1.24)}
     \Delta ^f_{\rm UV} 
     = \mu ^{-2\epsilon} \Delta _{\rm UV} \mu ^{2\epsilon}.
\end{equation}
The origin of the seemingly bizarre $\mu $-factors in \eq{(1.24)} is 
as follows. The \rop\ and the operator $\Delta _{\rm UV}$ should be 
applied to ``standard" dimensionally regularized Feynman integrals 
containing one factor $\mu ^{2\epsilon}$ per each loop. But the 
expression $e^l F(\varphi )$ generates integrals with one such factor 
lacking per each connected component---cf.\ the end of 
subsect.~\ref{ss1.3}. To remedy this, the powers of $\mu $ are 
introduced into \eq{(1.24)}, where it has been taken into account that 
$\Delta _{\rm UV}$ is non-trivial only on integrals with one connected 
component. It should be noted that in the context of the functional 
techniques $\Delta _{\rm UV}$ is systematically replaced by $\Delta 
^f_{\rm UV},$ so that the superscript $f$ can and will henceforth 
be omitted.

Using \eq{(1.23)}, we rewrite \eq{(1.22)} as:
\begin{equation}\label{(1.25)} 
     \hbox{eq.\eq{(1.22)}}
     = {\rm T} \prod_\alpha 
     \Delta _{\rm UV} \. ({\rm T} L^{k_\alpha}(\varphi )),
\end{equation}
where each T, as before, is the functional differential operator defined 
by \eq{(1.19)}.
Substituting this into \eq{(1.21)}, we finally obtain:
\begin{equation}\label{(1.26)}
     S_{\rm R}(\varphi ) = {\rm T} \exp[L_{\rm R}],
\end{equation}
where
\begin{equation}\label{(1.27)}
     L_{\rm R} = \Delta _{\rm UV}\. 
     \left( {\rm T}\,e^{L(\varphi)}-1 \right).
\end{equation}

Note that $\Delta _{\rm UV}\. (L(\varphi )) = L(\varphi ) $ (recall 
that $\Delta _{\rm UV}$ is a unit operation  on elementary vertices), 
therefore $L_{\rm R}$ can be represented as:
\begin{equation}\label{(1.28)}
     L_{\rm R} = L(\varphi ) + \hbox{``divergent\ UV-counterterms",}
\end{equation}
where
\begin{equation}\label{(1.29)}
     \hbox{``divergent\ UV-counterterms"} 
     = \Delta _{\rm UV}\.
         \left[ {\rm T}\,e^{L(\varphi)}-L(\varphi ) - 1
        \right]
\end{equation}
(note that each term in the  bracketed expression on the r.h.s.\ 
of \eq{(1.29)} that gives a non-zero result after application of 
$\Delta _{\rm UV}$ has at least one loop). So, indeed, the effect of 
the \rop\ is equivalent to adding some counterterms to the interaction 
Lagrangian. \Eq{(1.29)} provides a convenient explicit expression for 
them.

For clarity's sake, let us explain the algorithm encoded in 
\eq{(1.29)}:

(1) one writes down all the contributions to 
${\rm T}\,\exp[L(\varphi)] = \exp({1\over2}\delta \Delta^c \delta ) \exp[L(\varphi )];$

(2) then one discards all the terms except those that correspond to 
connected 1PI graphs with at least one loop;

(3) each of the terms left has the form:
\begin{equation}\label{(1.30)} 
     \mu ^{-2\epsilon} \int\,(\prod_i \,dp_i)(2\pi)^D 
          \,i\delta (\sum p_i) 
     FI(p_1\ldots p_k) \tphi (p_1)\ldots \tphi (p_k),
\end{equation}
where 
\begin{equation}\label{(1.31)}
     \tphi (p) = (2\pi)^{-D}\,\int\,e^{ipx}\, d^Dx\, \varphi (x),
\end{equation}
and $\delta (\sum p_i)$ expresses momentum conservation  while 
$FI(p_1\ldots p_k)$ is a connected 1PI Feynman integral constructed 
according to the rules described after \eq{(1.1)}. To $FI,$ one should 
apply $\Delta _{\rm UV}$ defined after \eq{(1.2)}, and $\Delta _{\rm 
UV}$ replaces $FI$ in \eq{(1.30)} by a polynomial of $p_i$ with 
coefficients that are divergent in the limit $D \to4.$ That is, 
$\Delta _{\rm UV}$ transforms \eq{(1.30)} into an integral over $x$ of 
local products of $\varphi (x)$ and its derivatives. (Note that in the 
case of the vacuum graphs ($k=0$) the $\delta $-function degenerates 
into an ill-defined factor $\delta (0),$ and a special axiom should 
fix a recipe for handling it. Such a recipe is completely 
non-interfering with what we are doing: the integral $FI\,$ is correctly 
defined even in this case for both the $R$- and $As$-operations to 
yield meaningful results when applied to it.)

The method for resolving combinatorics of 
the \rop\ that we have used is an extension to the more 
complicated case of the MS-scheme of a reasoning from subsect.~IV.1.3 
of \cite{zav}. In \cite{zav} the momentum subtraction scheme 
\cite{bog-shir} was used (so that only one term on the r.h.s.\ of 
\eq{(1.4)} was retained, namely, the one corresponding to the 
UV-subgraph containing {\em all} the lines of the graph $G$ which connect 
the vertices from the subset $v$), and it was assumed that the 
Lagrangian is normally ordered. As a result, in \cite{zav} the 
counterterms in \eq{(1.29)} were to be understood as normal products, 
while no such normal ordering is assumed in our case.

\subsection{\rop\ on Green functions}
\label{ss1.6}

Let us now turn to Green functions. To obtain the generating 
functional $G(J)$ for the Green functions one should add a source term 
to the Lagrangian:
\begin{equation}\label{(1.32)}
     L \to L + \varphi J,
\end{equation}
and evaluate the vacuum average of the resulting $S$-matrix.

The term $\varphi J$ generates a vertex that can be either isolated 
from, or connected by only one line with the rest of the graph. In 
either case the resulting graph is nullified by $\Delta _{\rm UV}$ 
unless it consists of a single vertex $\varphi J$ or does not contain  
such vertices at all. Formally:
\begin{equation}\label{(1.33)}
     \Delta _{\rm UV}\. \left( {\rm T}\, e^{L+\varphi J} \right) 
     = \Delta _{\rm UV} \. \left( {\rm T} e^L \right) 
     + \Delta _{\rm UV}\. (\varphi J)
     = \Delta _{\rm UV}\. \left( {\rm T} e^L \right) 
     + \varphi J.
\end{equation}
The last equation is due to the fact that $\Delta _{\rm UV}$ does not 
affect single vertices. Therefore, for the generating functional of 
the MS-renormalized Green functions one has:
\begin{equation}\label{(1.34)}
     G(J) \equiv \bold{R} \. <{\rm T}\,\exp[L(\varphi )+\varphi J]>_0
     = \left( {\rm T}\,\exp[L_{\rm R} + \varphi J] 
       \right)_{\varphi=0}.
\end{equation}
It may be helpful to transform \eq{(1.34)} to the form of a functional 
integral:
$$
     \left( {\rm T}\,\exp[\chi(\varphi)+\varphi J] \right)_{\varphi=0} 
     = \left( \exp({1\over2}\delta \Delta^c\delta) 
              \exp\left[ \chi \left( {\delta\over\delta J} \right) 
                   \right] 
              \exp\left[ \varphi J\right] 
       \right)
$$
$$
     = \exp\left[ \chi \left( {\delta\over\delta J} \right) 
           \right] 
       \exp\left({1\over2}J\Delta^c J \right) 
     = \exp\left[ \chi \left( {\delta\over\delta J} \right) 
            \right] \,
       \int\,d\varphi \exp[L_{\rm free} + \varphi J]
$$
\begin{equation}\label{(1.35)}
     = \int\,d\varphi \exp[L_{\rm free} + \chi(\varphi) + \varphi J],
\end{equation}
where 
\begin{equation}\label{(1.36)}
     L_{\rm free}(\varphi ) = -{1\over2}\varphi (\Delta^c)^{-1}\varphi 
\end{equation}
is the free classical action (multiplied by $i\mu ^{-2\epsilon}$) of 
the field $\varphi.$ Denoting by $L_{\rm tot}$ the full classical 
action (multiplied by $i\mu ^{-2\epsilon}$) of the model:
\begin{equation}\label{(1.37)}
     L_{\rm tot} = L_{\rm free} + L,
\end{equation}
and using \eq{(1.28)}, we have for \eq{(1.34)}:
\begin{equation}\label{(1.38)}
     G(J) = \int\,d\varphi\,  \exp[L_{\rm tot}(\varphi ) 
     + \hbox{``divergent\ UV-counterterms"} + \varphi J].
\end{equation}
This equation together with \eq{(1.29)} provides a convenient link 
between practical calculations of counterterms and the analysis of 
Green functions by the renormalization group method.

 
\section{Inversion of the \rop\ and the $\xi$-mapping}
\label{s2}

Since the effect of the \rop\ consists in adding to a graph $G$ a 
linear combination of graphs with a lesser number of loops, it turns 
out possible to construct an inversion of the \rop---in very much the 
same way as it is always possible to invert a triangle matrix with 
units on the main diagonal. We construct the inverted \rop, and to 
this end the so-called formalism of $\xi$-mapping is developed. This 
formalism proves to be a useful conceptual tool for studying various 
aspects of renormalization owing to the universality of the 
$\xi$-mapping, i.e.\ owing to the fact that it contains full information 
on the \rop\ in a given subtraction scheme for arbitrary models with a 
given field content. The main application of the inverted \rop\ will 
be the construction of Euclidean asymptotic expansion of Feynman 
diagrams in an explicitly convergent form.

After introducing some definitions in subsect.~\ref{ss2.1}, in 
subsect.~\ref{ss2.2} the $\xi$-mapping is defined. 
In subsects.~3.3--3.5 its 
inversion, the mapping $\xi^{-1}$ is proved to exist. A useful property 
of $\xi^{-1}$ is pointed out in subsect.~\ref{ss2.6}, and in 
subsect.~\ref{ss2.7} we briefly discuss how the structure of the 
renormalization group emerges from the point of view of the formalism 
of $\xi$-mappings. In subsect.~\ref{ss2.8} the inverted \rop\ is 
defined and an explicit algorithm for calculating the corresponding 
counterterms is derived.

\subsection{Local operators, local functionals, and Lagrangian 
functionals}\label{ss2.1}

Let $j_k(x)$ be a full linearly independent set of local products of 
$\varphi (x)$ and its derivatives. In momentum representation, one 
has:
$$
     \tj _{k}(q) = \int\,d^D p_1\ldots d^D p_m\, 
     (2\pi)^D  \mu ^{-2\epsilon} \delta 
     \left( q + \sum_{k=1}^m p_{k} \right) {\cal P}_{k}
$$
\begin{equation}\label{(2.1)} 
     \times \tphi (p_1)\ldots \tphi (p_{m}),
\end{equation}
with $\tphi (p)$ defined in \eq{(1.31)}; ${\cal P}_{k}$ is a 
polynomial of $p_1\ldots p_{m}.$ For convenience we assume that ${\cal 
P}_{k}$ are polynomials of the masses of the model and are allowed to 
contain factors depending only on $q=-\sum p_{k}$---or in terms of the 
coordinate representation, the local operators 
$j_{k}(x)$ are allowed to contain full derivatives in $x$.

We say that ${\cal L}$ is a {\em local functional} of fields if
\begin{equation}\label{(2.2)} 
     {\cal L} = \sum_n \, \int\, dq\, 
     {\cal L}_{n}(q) \tj _{n}(q),
\end{equation}
where ${\cal L}_{n}(q)$ are coefficient functions that are independent 
of the fields.

We call ${\cal L}$ defined in \eq{(2.2)} a {\em Lagrangian functional\/} if
\begin{equation}\label{(2.3)}
     {\cal L}_{n}(q) = g_{n}\delta (q),
\end{equation}
where $g_{n}$ are coupling constants independent of $q.$ For example, 
$L$ and $L_{\rm R}$ in \eqs{(1.14)}-{(1.16)} and \eq{(1.27)} 
are Lagrangian functionals.

\subsection{The $\xi$-mapping}
\label{ss2.2}

One can see that the proof of eqs.~\eq{(1.18)} and \eq{(1.26)} is 
valid not only for Lagrangian, but also for general local functionals 
${\cal L}:$
\begin{equation}\label{(2.4)} 
     \bold{R} \. {\rm T}\,\exp {\cal L} 
     = {\rm T}\,\exp \xi [{\cal L}],
\end{equation}
where
$$
     \xi [{\cal L}] 
     \equiv \Delta _{\rm UV}\. ({\rm T}\,\exp {\cal L} - 1).
$$
It is not difficult to see that $\xi [{\cal L}]$ is also a local 
functional:
\begin{equation}\label{(2.5)}
     \xi [{\cal L}] 
     = \sum_n \int\, dq \,\xi _{n}[{\cal L};q]\, \tj _{n}(q).
\end{equation}
In momentum representation its coefficient functions are formal series 
of the form:
$$
     \xi _{n}[{\cal L};q] 
     = \sum_{N=0}^\infty \sum_{n_1} \ldots \sum_{n_N} 
       \int\,dq_1 \ldots \int\,dq_N\, 
       \delta \left( q + \sum_{i=1}^N q_i \right)
$$
\begin{equation}\label{(2.6)}
     \times {\cal P}^{(N)}_{n;n_1\ldots n_N} (q_1\ldots q_N) 
     {\cal L}_{n_1}(q_1) \ldots {\cal L}_{n_N}(q_N),
\end{equation}
where ${\cal P}$ are polynomials in $q_i.$ The UV poles are contained 
in the coefficients of the polynomials ${\cal P}.$ In the coordinate 
representation, $\xi _{n}[{\cal L};x]$ are sums of local products of 
${\cal L}_i(x)$ and their derivatives.

The representation of $\xi [{\cal L}]$ in the form of \eq{(2.5)} and 
\eq{(2.6)} is in general not unique because a factor depending only on 
$q$ as a whole can be included either into $\xi _{n}[{\cal L};q]$ or 
into $\tj _{n}(q)$. But it can be made unique if one requires e.g.\ 
that 
${\cal P}$ from \eq{(2.6)} should not contain such factors. We assume 
that this condition is always fulfilled.

If ${\cal L}$ is a Lagrangian functional (cf.\ \eq{(2.3)}) then such is 
$\xi [{\cal L}]$ as well, as can be seen from \eq{(2.6)}.

So far we have done very little beyond rewriting the results of 
sect.~\ref{s1} in a new form. But this exercise is by no means trivial 
because it exhibits the fact that the $\xi$-mapping is universal in 
the sense that it depends only on the field content of the model but 
not on the interaction. We will see that the $\xi$-mapping is a 
convenient tool for studying UV renormalization of interaction 
Lagrangians as well as composite operators.

\subsection{Inversion of the $\xi$-mapping}
\label{ss2.3}

We are going to prove that there exists a mapping $\xi ^{-1}[{\cal 
L}]$ of the form of \eqs{(2.5)}-{(2.6)} with ${\cal P}$ 
replaced by some other polynomials $\bar{\cal P}$ satisfying the same 
restrictions. More precisely,
\begin{equation}\label{(2.7)} 
     \xi [\xi ^{-1}[{\cal L}]] = {\cal L} 
     \quad {\rm and} \quad
     \xi^{-1} [\xi [{\cal L}]] = {\cal L}
\end{equation}
for any local functional ${\cal L}.$ The equations \eq{(2.7)} allow 
one to write down the following inversion of \eq{(1.3)}:
\begin{equation}\label{(2.8)}
     \bold{R} \. {\rm T}\,\exp \xi ^{-1}[{\cal L}] 
     = {\rm T}\,\exp {\cal L}.
\end{equation}
This equation is one of the key results of the present paper.

\subsection{The structure of the $\xi$-mapping}
\label{ss2.4}

Let us study the structure of $\xi [{\cal L}].$ (Throughout this 
section we denote $\Delta \equiv \Delta _{\rm UV}.$)

The terms that are linear in ${\cal L}$ can be extracted from $\xi 
[{\cal L}]$ as follows:
\begin{equation}\label{(2.9)} 
     \xi [{\cal L}] 
     \equiv 
     \Delta \. ({\rm T}\, e^{\cal L}-1) 
     = \Delta \. ({\rm T}\, {\cal L}) 
     + \Delta \. {\rm T}\, \left( e^{\cal L}-{\cal L}-1 \right)
\end{equation}
Then
\begin{equation}\label{(2.10)} 
     \Delta \. ({\rm T}\, {\cal L}) 
     = \Delta \. {\rm T}\,
       \left[ \sum_m \int dq \,{\cal L}_{m}(q) \,\tj _{m}(q) \right]
     = \sum_m \int dq \,{\cal L}_{m}(q)\, 
       \Delta \. ({\rm T}\, \tj _{m}(q))
\end{equation}
and
\begin{equation}\label{(2.11)}
     \Delta \. ({\rm T}\, \tj _{m}(q)) = \sum_n z_{m,n}\, \tj _{n}(q).
\end{equation}
Indeed, $\tj _{m}(q)$ is the Fourier transform of a local products of 
fields so that the operation of $T$-product generates all possible 
tadpole graphs from  $\tj _{m}(q).$ $\Delta $ replaces the loop 
integrals corresponding to such diagrams by counterterms, and the 
result is a linear combination of local operators, which is what 
\eq{(2.11)} states. In any subtraction scheme
\begin{equation}\label{(2.12)}
     z_{n,n} = 1.
\end{equation}
In the MS-scheme, $z_{m,n}$ for $n\not=m$ are poles in $D-4$ with 
numeric coefficients that are independent of the dimensional 
parameters. (Recall that we include masses into the local operators 
$j.$)

Let us show that the matrix $z_{m,n}$ has an inverse. To this end we 
choose a special basis $\tj _{n}$ such that each $\tj _{n}$ is a 
monomial of fields and masses. Moreover, it is easy to see that the 
basis can be chosen to be ordered so that if the $n$-th operator is 
built of a lesser number of fields than the $n'$-th one (provided the 
dimensionalities in mass units and the numbers of full derivatives are 
the same) or if the dimensionality of the $n$-th operator is less than 
that of the $n'$-th one, then $n < n'.$ In such a basis, $m < n$ 
implies that $z_{m,n}= 0.$ Therefore, $z_{m,n}$ is a block-triangle 
matrix with units on the diagonal. It follows immediately that 
$z^{-1}$ exists and has the same block-triangle structure as $z,$ and 
the matrix elements of $z^{-1}$ are polynomials of non-zero elements 
of $z.$

\subsection{Existence of $\xi^{-1}$}
\label{ss2.5}

Now we can solve the equation
\begin{equation}\label{(2.13)}
     {\bar{\cal L}} = \xi [{\cal L}]
\end{equation}
with respect to ${\cal L}.$

Rewrite \eq{(2.7)} in the form of an equation for the sources:
\begin{equation}\label{(2.14)}
     {\bar{\cal L}}_{n}(q) = \xi _{n}[{\cal L};q],
\end{equation}
using the above information on the structure of $\xi ,$ extract the 
terms that are linear in ${\cal L}$ on the r.h.s.\ of \eq{(2.14)}:
\begin{equation}\label{(2.15)}
     {\bar{\cal L}}_{n}(q) = \sum_m {\cal L}_{m}(q)z_{m,n} 
     + {}_2 \xi _{n}[{\cal L};q].
\end{equation}
Multiplying \eq{(2.15)} by $z^{-1},$ we can rewrite it as:
\begin{equation}\label{(2.16)}
     {\cal L}_{n}(q) 
     = \sum_m {\bar{\cal L}}_{m}(q)z^{-1}_{m,n} 
     - \sum_m {}_2\xi _{m}[{\cal L};q]z^{-1}_{m,n}.
\end{equation}
Now the expression of ${\cal L}_{n}$ in terms of ${\bar{\cal L}}_{n}$ 
can be obtained by iterating \eq{(2.16)}, taking as a starting value 
the first term on the r.h.s.\ of \eq{(2.16)}. $k$ iterations allow one 
to obtain an expression of ${\cal L}_{n}$ in terms of ${\bar{\cal 
L}}_{n}$ to order $k$ in ${\bar{\cal L}}_{n}.$ Denoting the obtained 
solution of \eq{(2.16)} as
\begin{equation}\label{(2.17)}
     {\cal L} = \xi ^{-1}[{\bar{\cal L}}],
\end{equation}
one can easily see that the mapping $\xi ^{-1}$ has the properties 
described in subsect.~\ref{ss2.3}.

\subsection{A useful property of $\xi $ and $\xi ^{-1}.$}
\label{ss2.6}

If $J$ 
is a source for the field $\varphi $ then \eq{(1.33)} can be rewritten as:
\begin{equation}\label{(2.18)}
     \xi [{\cal L} + \varphi J] = \xi [{\cal L}] + \varphi J.
\end{equation}
It is not difficult to see that 
\begin{equation}\label{(2.19)}
     \xi ^{-1}[{\cal L} + \varphi J] 
     = \xi ^{-1}[{\cal L}] + \varphi J.
\end{equation}

\subsection{$\xi$-mapping and renormalization group}
\label{ss2.7}

It might be interesting to note that the above proof did not use any 
specific properties of the MS-scheme, though $\xi $ and $\xi ^{-1}$ do 
depend on the details of the \rop\ via ${\cal P}$ and ${\bar{\cal L}}$ 
in \eq{(2.6)}. It follows that the above formalism offers an 
alternative way of studying renormalization group problems.

For instance, let $\bold{R}_a$ and $\bold{R}_b$ be two \rop s 
differing by the choice of subtraction operators. Then:
\begin{equation}\label{(2.20)} 
     \bold{R} _a\. {\rm T}\, e^{\cal L} 
     = {\rm T}\, e^{\xi_a[{\cal L}]} 
     = \bold{R} _b \. {\rm T}\, e^{ \xi^{-1}_b [ \xi_a[ {\cal L} ] ] } 
     \equiv \bold{R} _b \. {\rm T}\, e^{ \xi^{-1}_{ab} [ {\cal L} ] }.
\end{equation}
One sees that $\xi   $ has the form of \eqs{(2.5)}-{(2.6)} 
with suitable polynomials ${\cal P}$ which---as can be easily 
understood---stay finite when the regularization is removed. Thus the 
structure of the renormalization group emerges very naturally%
\footnote{
It may be interesting to compare the above rather compact reasoning 
with the more cumbersome derivation of the renormalization group 
transformation given in \cite{bonneau} in the BPHZ framework.
}.

The Zimmermann identities for the Green functions with local operator 
insertions%
\footnote{
It should be noted that we use the term ``Zimmermann identities" in a 
slightly different sense than e.g.\ in \cite{bonneau}: we use it to 
denote the relations between the MS-renormalized and bare 
(unrenormalized) operators, without $\epsilon$-dependent factors like 
those considered in \cite{bonneau} for the purposes of studying 
anomalies.
}
can be easily obtained by performing suitable variations of 
\eq{(2.20)} with respect to the classical sources ${\cal L}_{n}$ 
entering ${\cal L}$ in \eq{(2.2)} (cf.\ sect.~\ref{s3} below).

\subsection{Practical calculations of $\bold{R} ^{-1}.$ }
\label{ss2.8}

Let us derive a diagram-by-diagram recipe for calculating the mapping 
$\xi ^{-1}.$ To this end it is convenient to introduce the inversion 
of the \rop, i.e.\ an operation $\bold{R} ^{-1}$ such that its 
structure is the same as that of the \rop\ (see \eq{(1.1)}) but the 
counterterms are evaluated according to a different rule in order to 
ensure that $\bold{R} ^{-1}\. \bold{R}  = 1$ and $\bold{R} \. \bold{R} 
^{-1}= 1.$

By definition, for the inverse \rop\ one would have:
\begin{equation}\label{(2.21)}
     \bold{R} ^{-1}\. {\rm T}\,e^{\cal L} 
     = {\rm T}\,\exp \xi ^{-1}[{\cal L}].
\end{equation}
There should exist the operator $\Delta ^{-1}$ related to $\bold{R} 
^{-1}$ in the same way as $\Delta $ to $\bold{R} $ (see \eq{(1.1)}; 
recall that throughout this section $\Delta \equiv \Delta _{\rm UV}$). 
Then:
\begin{equation}\label{(2.22)}
     \xi ^{-1}[{\cal L}] 
     \equiv  \Delta ^{-1}\. ({\rm T}\,e^{\cal L} - 1).
\end{equation}
From the second equation in \eq{(2.7)} the following condition on $\Delta 
^{-1}$ emerges:
\begin{equation}\label{(2.23)}
     {\cal L} 
     = \Delta ^{-1}\. [{\rm T}\,\exp \Delta \. 
       ({\rm T}\, e^{\cal L} - 1) - 1].
\end{equation}
\Eq{(2.23)} can be rewritten as:
$$
     \Delta ^{-1}\. ({\rm T}\, e^{\cal L} - 1 - {\cal L}) 
     + (\Delta ^{-1}\. {\cal L} - {\cal L})
$$
\begin{equation}\label{(2.24)}
     = - \Delta ^{-1} \. 
         [ {\rm T}\,\exp \Delta \. 
             ({\rm T}\, e^{\cal L} - 1) 
         - {\rm T}\, e^{\cal L} 
         ],
\end{equation}
where we have postulated linearity of $\Delta ^{-1}$. Note that the 
argument of $\Delta ^{-1}$ on the r.h.s.\ of \eq{(2.24)} is equal to 
$(\bold{R} -1)\. ({\rm T}\,e^{\cal L})$.

Assume that 
\begin{equation}\label{(2.25)}
     \Delta ^{-1}\. [{\cal L}] = {\cal L},
\end{equation}
which means that $\Delta ^{-1}$ is a unity operation on an elementary 
vertex. Then \eq{(2.24)} can be rewritten as:
\begin{equation}\label{(2.26)}
     \Delta ^{-1}\. ({\rm T}\, e^{\cal L} - 1 - {\cal L}) 
     = \Delta ^{-1}\. [(\bold{R}  - 1)\. {\rm T}\, e^{\cal L}].
\end{equation}
Therefore, for \eq{(2.23)} to hold, it is sufficient that the operator 
$\Delta ^{-1}$ has the following properties: linearity; the property 
\eq{(2.25)}; equality to 0 on unconnected and non-1PI diagrams; its 
action on 1PI diagrams should be defined by the following recursion:
\begin{equation}\label{(2.27)}
     \Delta ^{-1}\. [G] = - \Delta ^{-1}\. [\bold{R}  - 1]\. G.
\end{equation}
The r.h.s.\ should be understood as follows. $(\bold{R} -1)$ replaces 
$G$ by a sum of integrals obtained from $G$ by replacing one or more 
non-trivial UV-subgraphs by the corresponding UV-counterterm (see the 
definition of the \rop\ in subsect.~\ref{ss1.1}). Therefore, $\Delta 
^{-1}$ on the r.h.s.\ acts on a linear combination of Feynman integrals 
with divergent coefficients, all the integrals having one loop less as 
compared to $G$, while the divergent coefficients remain unaffected by 
$\Delta ^{-1}$ because of its linearity. On one loop integrals:
\begin{equation}\label{(2.28)}
     \Delta ^{-1}\. [g] 
     \equiv  - \Delta ^{-1}\. \Delta \. [g] 
     = - \Delta \. [g],
\end{equation}
and the recursion stops correctly.

\Eq{(2.27)} provides a most convenient recipe for explicit 
calculations. For example, in the scalar $g\varphi ^4$ model one has:
\begin{equation}\label{(2.29)}
\end{equation}
\vskip2.5in
and so on.

\thispagestyle{myheadings}\markright{}
\section{Renormalization of multilocal operator insertions}
\label{s3}

It is not difficult to generalize the above results to arbitrary Green 
functions of composite operators. The main point here is that it is 
simply sufficient to consider $T$-exponents of arbitrary local 
functionals ${\cal L}$ as defined in \eq{(2.2)} instead of Lagrangian 
functionals, and then derive expressions for specific composite 
(multi-) local operator insertions by performing suitable variations. 
In subsect.~\ref{ss3.1} we present some explicit formulae for the case 
when one wishes to express renormalized Green functions in terms of 
unrenormalized ones, and in subsect.~\ref{ss3.2} we consider the 
opposite case. Note that the case of one local operator insertion 
within the momentum subtraction scheme was first considered by 
Zimmermann \cite{zimm}.%
\footnote{
see also the remark in the footnote in the last paragraph of 
subsect.~\ref{ss2.7}.
\relax}

\subsection{Generalized Zimmermann identities}
\label{ss3.1}

Consider a Green function of arbitrary local operators (without loss 
of generality we take the local operators to belong to the basis 
introduced in subsect.~\ref{ss2.1}):
\begin{equation}\label{(3.1)}
     \bold{R} \. <{\rm T}\, \tj _1(q_1)\ldots \tj _{n}(q_{n}) e^L>_0,
\end{equation}
where $\tj _{k}$ are Fourier transformed local products of $\varphi 
(x)$ and its derivatives as defined in \eq{(2.1)}. Using the formalism 
introduced in the preceding section, one gets:
$$
     \hbox{eq.\eq{(3.1)} }
     = \left( {\delta \over \delta{\cal L}_{n_N} (p_N) } 
              \ldots 
              {\delta \over \delta{\cal L}_{n_1} (p_1) } 
              \, \bold{R} \. {\rm T}\, e^{\cal L} 
       \right)_{ {\cal L}=L,\,\,\varphi=0 } 
$$
$$
     = {\rm T}\, 
       \left( {\delta \over \delta{\cal L}_{n_N} (p_N) } 
              \ldots  
              {\delta \over \delta{\cal L}_{n_1} (p_1) } 
              e^{ \xi [ {\cal L} ] } 
       \right)_{ {\cal L}=L,\,\,\varphi=0 } 
$$
\begin{equation}\label{(3.2)}
     = {\rm T}\, 
       \left\{
          \left[ 
             e^{ -\xi [ {\cal L} ] }
             {\delta \over \delta{\cal L}_{n_N} (p_N) } 
             \ldots  
             {\delta \over \delta{\cal L}_{n_1} (p_1) } 
             e^{ \xi [ {\cal L} ] } 
          \right]_{{\cal L}=L}
          e^{ \xi [ L ] } 
       \right\}_{\varphi=0}.
\end{equation}
The expression in the square brackets on the r.h.s.\ of \eq{(3.2)} is a 
sum of products of terms like
\begin{equation}\label{(3.3)}
     \left[ 
         {\delta \over \delta{\cal L}_{i_n} (p_n) } 
         \ldots
         {\delta \over \delta{\cal L}_{i_1} (p_1) } 
         \xi [{\cal L}] 
      \right] 
     = \Delta \. 
     \left[ 
         {\rm T} \, \tj_{i_1}(p ) \ldots \tj_{i_n}(p_{n}) \exp\,L 
     \right] .
\end{equation}
It is not difficult to see that \eq{(3.3)} is a local functional which 
can be expanded in $\tj $ as:
\begin{equation}\label{(3.4)}
     \hbox{eq.\eq{(3.3)}} 
     = \sum_m \tj _{m} ( p_1 + \ldots + p_{n} ) 
        \,{\tilde Z}^{(n)} _{i_1\ldots i_n;m} (p_1\ldots p_{n}),
\end{equation}
where ${\tilde Z}$ are polynomials in $p$ and in the parameters of 
$L.$ $Z$ is independent of $p$ if the dimensionality of $\tj $ in 
units of mass is greater than that of the product $\tj _{i_1}\ldots 
\tj _{i_n}.$ Moreover, since we include full derivatives into $j,$ $Z$ 
can not contain factors depending only on $p_1 + \ldots + p_{n};$ in 
particular, ${\tilde Z}^{(1)}_n(p)$ are independent of $p.$

Examples will help to understand the above. Denoting \eq{(3.3)} as 
$\Delta (j_1 \ldots j_{n}),$ one has ($L_{\rm R}$ was defined in 
\eq{(1.27)}):
\begin{equation}\label{(3.5')}
     \bold{R} \. <{\rm T}\, j_1e^L>_0 
     = ({\rm T}\, \Delta (j_1)\,\exp [L_{\rm R}]
       )_{\varphi=0}, 
\end{equation}

\begin{equation}\label{(3.52)}
     \bold{R} \. <{\rm T}\, j_1 j_2 e^L>_0 
     = ({\rm T}\, [ \Delta (j_1) \Delta (j_2) 
                  + \Delta (j_1 j_2)
                  ] \exp [ L_{\rm R} ]
       )_{\varphi=0},
\end{equation}

$$
     \bold{R} \. <{\rm T}\, j_{1}j_{2}j_{3}\,e^L>_0 
     = ( {\rm T} \, [ \Delta (j_{1}) \Delta (j_{2}) \Delta (j_{3}) 
                    + \Delta (j_{1}j_{2}) [ \check\Delta (j_{3})
$$

\begin{equation}\label{(3.53)}
     + \check\Delta ( j_{1}j_{3} ) \check\Delta ( j_{2} ) 
     + \check\Delta ( j_{2}j_{3} ) \check\Delta ( j_{1} ) 
     + \check\Delta ( j_{1}j_{2}j_{3} ) ] 
     \exp [ L_{\rm R} ] )_{\varphi=0}.
\end{equation}

Since $\check\Delta (\varphi )=\varphi $ and $\check\Delta (\varphi 
j_{1}\ldots j_{k})=0$ (this is because it is impossible to form a 1PI 
graph with one ``$\varphi $-insertion"), one can see that for 
$j=\varphi $ the above formulae degenerate into
\begin{equation}\label{(3.6)}
     \bold{R} \. <{\rm T}\, \tphi (q_1) \ldots \tphi (q_{n}) e^L>_0 
     = ( {\rm T} \, \tphi (q_1) \ldots \tphi (q_{n}) 
         e^{ L_{\rm R}(\varphi) } 
       )_{\varphi=0},
\end{equation}
which agrees with \eq{(1.34)}, as expected.

Using \eq{(3.4)}, one can represent \eq{(3.5')} as:
\begin{equation}\label{(3.7')}
     \bold{R} \. ( {\rm T}\, \tj _1(p) e^L ) 
     = \sum_m { \tilde Z }^{(1)}_{1;m} 
     ( {\rm T}\, \tj _{m}(p) \exp [ L_{\rm R} ] )
\end{equation}
(identities of this kind were first obtained by Zimmermann \cite{zimm}),
$$
     \bold{R} \. ( {\rm T}\, \tj _1(p_1)\tj _2(p_2)e^L )
     = \sum_m {\tilde Z}^{(2)}_{1,2;m} (p_1,p_2) 
       ( {\rm T} \, \tj _{m} ( p_1 + p_2 ) \exp[L_{\rm R}] )
$$
\begin{equation}\label{(3.72)}
       + \sum_m {\tilde Z}^{(1)}_{1;m} 
       \sum_n {\tilde Z}^{(1)}_{1;n} 
       ( {\rm T}\, \tj _{m}(p_1) \tj _{n}(p_2) \exp[L_{\rm R}] ),
\end{equation}
etc. Note that all the sums in \eq{(3.7')} and \eq{(3.72)} are finite. 
Identities like \eq{(3.72)} and their generalizations can be called 
(generalized) Zimmermann identities for multilocal operator 
insertions.

\subsection{Inverted Zimmermann identities}
\label{ss3.2}

The equations derived in the preceding subsection allow one to express 
renormalized Green functions in terms of unrenormalized ones. Using 
the inversion of the \rop\ obtained in sect.~\ref{s2}, one can easily 
do the opposite, i.e.\ reexpress {\em un}renorma\-lized quantities in 
terms of renormalized ones. Thus, instead of 
\eqs{(3.1)}-{(3.2)} one has:
$$
     <{\rm T}\, \tj _1(q_1)\ldots \tj _{n}(q_{n}) e^L>_0
$$
$$
     = \left( 
         { \delta \over \delta{\cal L}_{n_N} (p_N) } 
         \ldots  
         { \delta \over \delta{\cal L}_{n_1} (p_1) } 
         {\rm T}\, e^{\cal L} 
       \right)_{{\cal L}=L,\,\,\varphi=0}
$$
\begin{equation}\label{(3.8)}
     = \bold{R} \. {\rm T}\, 
       \left\{ 
          \left[ 
             e^{-\bar\xi [ {\cal L} ] } 
             {\delta \over \delta{\cal L}_{n_N} (p_N) } 
             \ldots  
             {\delta \over \delta{\cal L}_{n_1} (p_1) } 
             e^{ \bar\xi [ {\cal L} ] } 
          \right]_{{\cal L}=L} 
          e^{ \bar\xi [ L ] } 
       \right\}_{\varphi=0}, 
\end{equation}
and instead of \eq{(3.7')} one has:
\begin{equation}\label{(3.9')}
     {\rm T}\, \tj _1(p)\, e^L 
     = \sum_m \tilde{\tilde Z}^{(1)}_{1;m} \, 
        \bold{R} \. ( {\rm T}\, \tj _{m}(p) \,\exp\,\xi ^{-1}[L] )
\end{equation}
(note that 
$\tilde{\tilde Z}^{(1)}
=[{\tilde Z}^{(1)}]^{-1}
$), 
$$
     {\rm T}\, \tj _1(p_1)\tj _2(p_2)e^L 
     = \sum_m \tilde{\tilde Z}^{(2)}_{1,2;m}(p_1,p_2) 
         \bold{R} \. ( {\rm T}\, \tj _{m} (p_1 + p_2) 
                       \,\exp\,\xi ^{-1}[L]
                     )
$$
\begin{equation}\label{(3.92)}
     + \sum_m \tilde{\tilde Z} ^{(1)}_{1;m} 
       \sum_n \tilde{\tilde Z} ^{(1)}_{1;n} 
         \bold{R} \. ( {\rm T}\, \tj _{m}(p_1) \tj _{n}(p_2)
                        \,\exp\xi ^{-1}[L]
                     ),
\end{equation}
(one can check that 
$ {\tilde{\tilde Z}}^{(2)}_{1,2;m}(p_1,p_2) 
= - \sum_l {\tilde Z}^{(2)}_{1,2;l}(p_1,p_2) 
           {\tilde{\tilde Z}}^{(1)}_{l;m}
$). And so on.

In a similar manner one can relate expressions renormalized in 
different renormalization schemes,  cf.\ subsect.~\ref{ss2.7}.

\vskip2cm
\centerline{{\large\bf
$AS$-OPERATION FOR MULTILOOP INTEGRALS.
}}

\section{Heavy mass expansions}
\label{s4}

We now turn to asymptotic expansions of Feynman diagrams. In this 
section we consider the case when the set of the heavy parameters 
contains only masses, which corresponds to studying low-energy 
effective Lagrangians to all orders in the inverse heavy mass. The 
reasoning in the more general case is very much the same, so that in 
the next section we will be able to concentrate only on specific 
complications due to large external momenta. 

In subsect.~\ref{ss4.1} the \asop\ for unrenormalized (or 
UV-convergent) integrals is defined as an integrated version of the 
\asop\ for products of singular functions introduced in \cite{partI}, 
in subsect.~\ref{ss4.2} the combinatorial structure of the 
corresponding subgraphs is studied. In subsect.~\ref{ss4.3} the 
$\delta$-functions corresponding to the counterterms for the 
IR-subgraphs are integrated out and in subsect.~\ref{ss4.4} the final 
form of the \asop\ for 1PI integrals is presented. In 
subsect.~\ref{ss4.5} the obtained expression for the \asop\ is extended 
to non-1PI integrals, in subsect.~\ref{ss4.6} its exponentiation is 
proved, and in subsect.~\ref{ss4.7} it is transformed to an explicitly 
convergent form using the inversion of the \rop. In subsects.~4.8 and 4.9 
the results are extended to UV-renormalized diagrams and models, 
respectively. Explicit expressions for low-energy effective 
Lagrangians are presented to all orders of the inverse heavy mass. 
Subsect.~\ref{ss4.10} contains a simple example.

\subsection{\asop\ for integrals: definition}
\label{ss4.1}

Let us introduce notations and formulate the problem that we are going 
to consider here. Our notations on the whole will be consistent with 
subsect.~\ref{ss2.1} of \cite{partI}.

Let $\Gamma $ be an $l$-loop 1PI Feynman diagram, and let 
$p=(p_1\ldots p_l)$ denote its integration momenta. We assume in this 
section that $\Gamma $ depends on the light and heavy masses $m$ and 
$M,$ and that all the external momenta of $\Gamma $ are light (and 
denoted as $k$). All $M$ are non-zero and independent of the expansion 
parameter denoted as $\kappa ,$ while $m,k=O(\kappa )$ and some masses 
$m$ may be equal to zero. Unlike the abstract analysis of 
\cite{partI}, here it will be convenient to show the dependence on $m,
$ $k$ and $M$ explicitly.

Thus, the unrenormalized integrand is denoted as $\Gamma (p,m,k,M).$ 
We assume that $\Gamma $ is UV-convergent, i.e.\ $\Gamma (p,\ldots )$ 
is absolutely integrable over $p$ in infinite limits. Denote the 
integrated diagram as:
\begin{equation}\label{(4.1)}
     \Gamma (k,m,M) \mathop=^{\rm def} \,\int\,dp\, \Gamma (p,k,m,M).
\end{equation}
In \cite{partI} we introduced the \asop\ for products of singular 
functions like $\Gamma (p,k,m,M)$ which yields asymptotic expansions 
of such products with respect to the light parameters in the sense of 
the distribution theory. It is natural to expect (for more details see 
subsect.~\ref{ss4.8}; a fuller justification is given in \cite{IV}) 
that if \eq{(4.1)} is UV-convergent, then the 
asymptotic expansion of $\Gamma (k,m,M)$ can be obtained by expanding 
its integrand $\Gamma (p,k,m,M)$ using the \asop\ of \cite{partI} and 
then performing termwise integration of the resulting series over $p.$

Let us introduce an operation that acts on {\em integrated} Feynman 
diagrams in accordance with the above recipe. This new operation will 
be referred to as {\em the \asop\ for integrals.} To distinguish the 
two types of \asop s, one could use e.g. the notations $\bold{As}_{\rm 
prod}$ and $\bold{As}_{\rm int}.$ But as it is normally clear whether 
the \asop\ acts on a product of singular functions or on an integral, 
we will use the same notation $\bold{As}$ without subscripts in both 
cases. Formally, one has:
\begin{equation}\label{(4.2)} 
     \bold{As} \. \Gamma (k,m,M) 
     \mathop=^{\rm def} \,
     \int\, dp \,\bold{As} \. \Gamma (p,k,m,M).
\end{equation}

Let us exhibit the structure of the \asop\ on the r.h.s.\ of \eq{(4.2)}. 
The general expression for it was given in 
\cite{partI}. Specifying it to the case considered here, one obtains:
$$
     \bold{As} \. \Gamma (p,k,m,M) \kern9cm
$$
\begin{equation}\label{(4.3)}  
     = \bold{T}_{m,k} \. \Gamma (p,k,m,M) 
     + \sum_\gamma [ \bold{E} \. \gamma (p,k,m)]
          \times [\bold{T}_{m.k} \. {\Gamma\backslash\gamma }(p,k,m,M)
                 ],
\end{equation}
where the notations used are as follows.

The operation $\bold{T}_{m,k}$ performs the Taylor expansion in powers 
of the light parameters. Summation runs over all IR-subgraphs of 
$\Gamma .$ Specifying to the present case the definitions of 
subsects.~9.2--3 of \cite{partI}, one sees that an IR-subgraph $\gamma $ 
can be described as follows: $(i)$ the set of lines of $\gamma $ is a 
full subset of the light lines of $\Gamma ,$ i.e.\ if one puts to zero 
the momenta flowing through all the lines of $\gamma $ together with 
the light external momenta $k,$ then no light line which does not 
belong to $\gamma $ should have the momentum flowing through it 
nullified owing to momentum conservation (cf.\ 
\cite{partI}); $(ii)$ any vertex of $\Gamma $ whose incident lines all 
belong to $\gamma ,$ is included into $\gamma .$ The product of the 
lines and vertices of $\gamma $ is independent of $M$ and is denoted 
as $\gamma (p,k,m).$ 

The product $\Gamma \backslash\gamma $ in \eq{(4.3)} comprises all the 
factors from $\Gamma $ that do not belong to $\gamma .$ Graphically, 
$\Gamma \backslash\gamma $ is obtained from $\Gamma $ by deleting all 
the lines and vertices of $\gamma .$

The operation $\bold{E}$ (expressions for it were derived in 
 \cite{partI}) applied to an IR subgraph returns a 
linear combination of counterterms that are proportional to 
$\delta$-functions of $p.$ Our first aim here is to integrate out 
those $\delta$-functions explicitly.

\subsection{Structure of $\Gamma \backslash\gamma .$}
\label{ss4.2}

It is not difficult to verify that in the case under consideration (no 
heavy external momenta) the IR-subgraphs are uniquely characterized by 
the properties of their complements. Indeed, denote the connected 
components of $\Gamma\backslash\gamma $ as $h_i .$ These subgraphs 
(which will be referred to as {\em heavy knots}) have the following 
properties:

$(i)$ since $\gamma $ consists of only light lines, each $h_{i}$ has 
only light external lines while all the heavy lines of $\Gamma $ are 
hidden within all $h_{i};$

$(ii)$ since $\gamma $ is full, none of $h_{i}$ can have a light line 
whose momentum vanishes due to momentum conservation when one puts to 
zero all the external momenta of $h_{i}.$ This is an analytical 
formulation. Graphically, it implies that each $h_{i}$ is ``1PI with 
respect to light lines" which means that it cannot be divided into two 
disconnected graphs (which may consist of a single vertex) by cutting 
any one of its light lines.

Note that a heavy knot $h_{i}$ may consist of a single heavy line 
(including both vertices to which it is attached).

One has:
\begin{equation}\label{(4.4)}
     \Gamma \backslash\gamma = \prod_i h_{i}.
\end{equation}
In the final form of the \asop\ for integrals, the heavy knots will 
play the same role as UV-subgraphs in the expression \eq{(1.1)} for the 
\rop. 

{\bf Example}. 
Consider the diagram $\Gamma $ as shown in Fig.~1a. Fat lines 
correspond to propagators of heavy particles, the rest to light 
particles. In Figs.~1b--1d various IR subgraphs are shown with dashed 
lines. Heavy knots in each case are clearly visible as connected 
components built of solid---normal and fat---lines.

\subsection{Integrating $\delta$-functions}
\label{ss4.3}

Before substituting explicit expressions for $\bold{E} \. \gamma $ 
into \eq{(4.3)}, the dependence of each term in \eq{(4.3)} on $p$ 
should be studied. One can see that $\gamma$  does not depend on those 
components of $p$ that correspond to the loops of $\Gamma 
\backslash\gamma .$ Let us denote the collection of such momenta as 
$p_{\Gamma\backslash\gamma,{\rm int}}.$ Furthermore, each loop of 
$\Gamma \backslash\gamma $ belongs to only one heavy knot $h_{i}$ from 
the decomposition \eq{(4.4)}, so that:
\begin{equation}\label{(4.5)}
     p_{\Gamma\backslash\gamma,{\rm int}} = \{p_{i,{\rm int}} \}_{i}.
\end{equation}
The remaining momentum components of $p$ form exactly the proper 
variables of $\gamma $ introduced in  
\cite{partI} and denoted as $p_{\gamma },$ so that
\begin{equation}\label{(4.6)}
     p = (p_{\gamma },p_{\Gamma\backslash\gamma,{\rm int}}).
\end{equation}
Note that $\Gamma \backslash\gamma $ is independent of those 
components of $p_{\gamma }$ that correspond to the loops of $\gamma .$ 
Denote as $p_{\Gamma\backslash\gamma,{\rm ext}}$ those components of 
$\gamma $ on which $\Gamma \backslash\gamma $ does depend.

Substituting \eq{(4.3)} into \eq{(4.2)} and using the explicit 
expressions for $\bold{E} \. \gamma $ derived in 
\cite{partI}, we can rewrite \eq{(4.2)} as:
$$
     \bold{As} \. \Gamma (k,m,M) = \bold{T}_{m.k} \. \Gamma (k,m,M)\kern5cm
$$
$$
     + \sum_{\gamma}\sum_{\alpha}
       \bigl\{ 
          \left[ \,\int\, dp'_{\gamma } 
                 {\cal P}_{\gamma,\alpha}(p'_{\gamma })\, 
                 \gamma (p'_{\gamma },k,m) 
          \right]\,
          \int\, dp_{\gamma }\,
          \int\, dp_{\Gamma\backslash\gamma,{\rm int}} 
          \delta _{\gamma ,\alpha} (p_{\gamma })
$$
\begin{equation}\label{(4.7)}
     \times \bold{T}_{m,k} \. 
         \left[ 
            \Gamma\backslash\gamma
             (p_{\Gamma\backslash\gamma,{\rm ext}},
              p_{\Gamma\backslash\gamma,{\rm int}},k,m,M)
         \right]
       \bigr\},
\end{equation}
The polynomials ${\cal P}_{\alpha }$ and the $\delta$-functions 
$\delta _{\alpha }$ in \eq{(4.7)} form full dual sets (cf. 
 \cite{partI}) so that the sum over $\alpha$ 
represents the Taylor expansion:
\begin{equation}\label{(4.8)}
     \sum_{\alpha} 
         {\cal P}_{\gamma ,\alpha }(p'_{\gamma }) 
         \,\int dp_{\gamma } \delta _{\gamma ,\alpha }(p_{\gamma })
         \times f(p_{\gamma }) 
     \equiv \bold{T}_{p'_{\gamma}} \. f(p'_{\gamma }).
\end{equation}
(It does not matter that the expression that plays the role of $f$ in 
\eq{(4.7)} depends not on all $p_{\gamma }$ but only on some of the 
components of $p_{\gamma }$, namely, on $p_{\Gamma\backslash\gamma,
{\rm ext}}$.) Using \eq{(4.8)} and renaming $p'_{\gamma,{\rm ext}} \to 
p_{\gamma,{\rm ext}}$, we get:
$$
     \bold{As} \. \Gamma (k,m,M) = \bold{T}_{m,k} \. \Gamma (k,m,M) \kern4cm
$$
\begin{equation}\label{(4.9)}
     + \sum_\gamma \left\{ \,\int\, dp \gamma (p_{\gamma },k,m)
     \times \bold{T}_{m,k}\.\bold{T}_{p_{\Gamma\backslash\gamma,
                                         {\rm ext}}} \. 
     \left[ \Gamma\backslash\gamma (p_{\Gamma\backslash\gamma,
                                      {\rm ext}},
                                    p_{\Gamma\backslash\gamma,
                                      {\rm int}},
                                    k,m,M)
     \right]
     \right\}.
\end{equation}
Now recall the representation of $\Gamma \backslash\gamma $ in terms 
of heavy knots \eq{(4.4)}, \eq{(4.5)}. It is easy to see that the last 
line in \eq{(4.9)} can be rewritten as
\begin{equation}\label{(4.10)}
     \prod_i \tau \. h_{i}(p_{i,{\rm int}}, p_{i,{\rm ext}},k,m,M),
\end{equation}
where $p_{i,{\rm ext}}$ comprises those components of 
$p_{\Gamma\backslash\gamma,{\rm ext}}$ on which the heavy knot $h_{i}$ 
depends. The operation $\tau $ in \eq{(4.10)} Taylor-expands in $m$ 
and in all the momenta that are external with respect to the subgraph 
$h_{i},$ i.e.\ both $p_{i,{\rm ext}}$ and the corresponding components 
of $k.$ It should be stressed that each factor in \eq{(4.10)} is 
independent of the rest of the original diagram $\Gamma .$ Therefore, 
we arrive at the following expression of the \asop\ for integrated 
Feynman diagrams:
\begin{equation}\label{(4.11)}
     \bold{As} \. \Gamma  
     = \tau  \. \Gamma 
     + \sum_{\{h_i\}} \left( \prod_i \tau \. h_{i} \right) 
       \times [ \Gamma / \prod_i h_{i} ].
\end{equation}
The meaning of this formula is as follows. One enumerates all sets of 
pairwise non-intersecting heavy knots $h_{i}$ and contracts each heavy 
knot $h_{i}$ to a point, replacing it by a formal series in its 
external momenta which is obtained by Taylor-expanding the 
corresponding Feynman integral.

To avoid confusion, it should be understood that the IR-counterterms 
$\bold{E} \. \gamma $ in \eq{(4.3)} have transformed into the 
expression in the square brackets in \eq{(4.11)}, while the 
``counterterms" $\tau  \. h_{i}$ in \eq{(4.11)} have originated from 
the expression $\Gamma \backslash\gamma $ in \eq{(4.3)}. In other 
words, ``counterterms" and ``non-counterterms" have changed places in 
the transition from \eq{(4.3)} to \eq{(4.11)}.

\subsection{\asop\ for integrals: the final form}
\label{ss4.4}

\Eq{(4.11)} can be rewritten in the following form which is analogous 
to the representation \eq{(1.1)} of the \rop:
\begin{equation}\label{(4.12)}
     \bold{As} \. \Gamma 
     = \sum_{\{h_i\}} \left( \prod_i\Delta_{\rm as} \. h_{i} \right)
       \times [\Gamma  / \prod_i h_{i}]_{M=\infty}.
\end{equation}
The summation here runs over all sets of pairwise non-intersecting 
subgraphs $h_{i}$ of $\Gamma $ such that each vertex of $\Gamma $ 
belongs to one of $h_{i}.$ (A subgraph $h$ is defined as a subset of 
vertices of $\Gamma $ together with some lines of $\Gamma $ which are 
connected by both ends to the vertices included into $h.$ Subgraphs 
are non-intersecting if they have no common vertices and, 
consequently, no common lines.) One of such sets consists of a single 
subgraph $h=\Gamma $ which corresponds to the first term $\tau \. 
\Gamma $ on the r.h.s. of \eq{(4.11)}. The operation $\Delta_{\rm as}$ 
coincides with $\tau $ on heavy knots, is a unity operation on single 
vertices without loops, and returns zero otherwise. Setting $M=\infty 
$ is a formal expression of the fact that one should discard from the 
r.h.s. those terms in which some heavy lines remained outside all 
$h_{i}$ so that \eq{(4.11)} could be restored from \eq{(4.12)}.

There is no need to introduce special rules for heavy knots containing 
tadpoles made of light lines (cf.\ Fig.~2), our formalism takes them 
into account correctly. For practical purposes it is sufficient to 
note that such heavy knots are nullified by the operation $\tau $ due 
to the well-known property of the dimensional regularization to 
nullify momentum integrals without dimensional parameters.

The combinatorial resemblance of \eq{(4.12)} and the expression 
\eq{(1.1)} for the \rop\ is nearly complete except for the minor 
difference described above. However, this difference can be easily 
taken into account to make the reasoning of sects.~1--3 to be applicable 
to the \asop. But first expansions of diagrams $\Gamma $ other than 
1PI should be considered.

\subsection{\asop\ for non-1PI diagrams}
\label{ss4.5}

We have derived explicit expressions for the operation $\bold{As}$ on 
1PI integrals. Remarkably, one can expand arbitrary---not necessarily 
1PI---integrals using the same expression \eq{(4.12)}.

For the operation $\bold{As}$ to be applicable to any graph $\Gamma ,$ 
its structure should satisfy certain {\it a priori} requirements. Let 
us first enumerate them.

If $\Gamma $ is disconnected, i.e.\ there are two subgraphs $\Gamma 
_{1}$ and $\Gamma _{2}$ such that $\Gamma =\Gamma _{1}\times \Gamma 
_{2},$ then the expansion of $\Gamma $ is a product of the expansions 
of $\Gamma _{1}$ and $\Gamma _{2},$ i.e.
\begin{equation}\label{(4.13)}
     \bold{As} \. (\Gamma _{1}\times \Gamma _{2}) 
     = (\bold{As} \. \Gamma _{1})\times (\bold{As} \. \Gamma _{2}).
\end{equation}

If $\Gamma $ falls into two parts connected by only one light line 
(the corresponding propagator is denoted as $D_{l}$), i.e.\ $\Gamma 
=\Gamma _{1}\times D_{l}\times \Gamma _{2},$ then
\begin{equation}\label{(4.14)}
     \bold{As} \. (\Gamma _{1}\times D_{l}\times \Gamma _{2}) 
     = ( \bold{As} \. \Gamma _{1}) \times D_{l} 
         \times (\bold{As} \. \Gamma _{2} ).
\end{equation}
This is due to the fact that both the mass of this line and the 
momentum flowing through it (which is a combination of the light 
external momenta of $\Gamma $) are expansion parameters so that 
$D_{l}$ is proportional to a negative power of $\kappa $ and need not 
be expanded.

If $\Gamma $ consists of two subgraphs connected by one heavy line 
(the corresponding propagator is denoted as $D_{h}$), i.e.\ $\Gamma 
=\Gamma _{1}\times D_{h}\times \Gamma _{2},$ then
\begin{equation}\label{(4.15)}
     \bold{As} \. (\Gamma _{1}\times D_{h}\times \Gamma _{2}) 
     =       (\bold{As}    \. \Gamma _{1} )
       \times (\bold{T}_{k} \. D_{h}       )
       \times (\bold{As}    \. \Gamma _{2} ).
\end{equation}
(Recall that $\bold{T}_{k}$ Taylor-expands in $k,$ and that the 
momentum flowing through $D_{h}$ is a combination of the light 
external momenta of $\Gamma .$)

Now, note that the definition \eq{(4.12)} does not use 1PI-ness of 
$\Gamma .$ Therefore, let us define the operation $\bold{As}$ by 
\eq{(4.12)} on any graph $\Gamma .$ Let us prove that the operation 
thus defined satisfies the above {\it a priori} criteria 
\eqs{(4.13)}-{(4.15)}.

\Eq{(4.13)} is true because each $h_{i}$ from \eq{(4.12)} lies 
strictly within either $\Gamma _{1}$ or $\Gamma _{2},$ so that each 
partition of $\Gamma $ consists of a partition of $\Gamma _{1}$ and a 
partition of $\Gamma _{2},$ and a sum over all partitions of $\Gamma $ 
is equivalent to a sum over partitions of $\Gamma _{1}$ times a sum 
over partitions of $\Gamma _{2}.$

The same argument proves \eq{(4.14)} because $D_{l}$ cannot belong to 
any of $h_{i}$ without making it 1PI with respect to light lines, 
which implies that each $h_{i}$ with non-zero $\Delta _{as}$ lies 
strictly within either $\Gamma _{1}$ or $\Gamma _{2}.$

To prove \eq{(4.15)} we note that $D_{h}$ must always belong to one of 
the $h_{i}.$ Denote this heavy knot as $h.$ Then $h=h'\times D_{h}\times 
h''$ (where $h'$ and $h''$ each consist of at least one vertex) and
\begin{equation}\label{(4.16)}
     \Delta _{as} \. (h' \times D_{h}\times h'') 
     = (\Delta _{as} \. h') 
       \times 
       (\bold{T}_{k} \. D_{h})\times (\Delta _{as} \. h'').
\end{equation}
Then one applies the same argument as above.

An induction with respect to the number of 1PI components of $\Gamma $ 
enables one to conclude that \eq{(4.12)} remains correct on all 
Feynman diagrams.

\subsection{Exponentiation of $\Delta _{\rm as}.$}
\label{ss4.6}

Now we are in a position to apply the combinatorial results of 
sects.~1--3 to \eq{(4.12)}.

Let $\varphi $ and $\Phi $ be the fields of the light and heavy 
particles that have the masses $m$ and $M,$ respectively. Let $L = 
L(\varphi ,\Phi )$ be an interaction Lagrangian. Consider the 
generating functional of Green functions of 
the light fields $<{\rm T} \exp [ L(\varphi ,
\Phi ) + \varphi J ]>_{0}.$ It is assumed that all the external 
momenta of the Green functions are light, i.e.\ $O(\kappa).$ We ignore 
the UV divergences for a while, because we are now interested only in 
the formal global structure of the operation $\bold{As};$ one can 
simply assume that there are no UV divergences at all in the model 
described by $L.$

As was noted above, the combinatorial structures of the $As$- and 
\rop s are very similar. The similarity goes far enough to allow one 
to write down the following analogue of \eq{(1.32)}:
\begin{equation}\label{(4.17)}
     \bold{As} \,\. <{\rm T} \exp[L(\varphi ,\Phi )+\varphi J]>_{0} 
     \;=\; <{\rm T} \exp [ L_{\rm eff,bare}(\varphi ) 
                     + \varphi J ]>_{0},
\end{equation}
where
\begin{equation}\label{(4.18)}
     L_{\rm eff,bare}(\varphi ) 
     = \Delta _{\rm as} \. \left[ {\rm T} e^{L(\varphi,\Phi)} 
                            - 1
                            \right]_{\Phi=0}.
\end{equation}
Setting $\Phi $ to zero in \eq{(4.18)} is formally equivalent to 
setting $M = \infty $ in \eq{(4.12)} and is an expression of the fact 
that each heavy line in \eq{(4.11)} must belong to one of the heavy 
knots. 

The algorithm of \eq{(4.18)} is as follows:

$(i)$ all the diagrams contributing to ${\rm T} \exp L(\varphi ,\Phi )$ 
are evaluated;

$(ii)$ $\Delta _{\rm as}$ nullifies all the terms except those which 
correspond to connected diagrams that are 1PI with respect to light 
lines;

$(iii)$ on single vertices, $\Delta _{\rm as}$ acts as a unit operation, 
i.e.\ 
\begin{equation}
\Delta _{\rm as} \. [L(\varphi ,\Phi )]_{\Phi=0} 
= [L(\varphi ,\Phi )]_{\Phi=0};
\end{equation}

$(iv)$ the remaining terms are exactly the heavy knots and can be 
represented as \eq{(1.29)}, where $FI$ also depends on $m$ and $M.$ 
$\Delta _{\rm as}$ performs the formal Taylor expansion of $FI$ with 
respect to $m$ and all the external momenta of such terms, i.e.\ 
$p_{i}$ in \eq{(1.29)}.

Note that the recipe for $\Delta _{\rm as}$ allows a remarkably 
``fool-proof" reformulation (which may be useful e.g.\ 
in writing computer 
programs): it is sufficient to apply the $\tau $-operation to {\em 
all} connected terms generated by the $T$-exponent. Then the terms 
that should be discarded will either develop singularities 1/0, or be 
nullified owing to the property of the dimensional regularization to 
nullify integrals without dimensional parameters like
$$
   \,\int\, d^Dp \, p^{-2a} = 0.
$$
It is no less remarkable that this ``fool-proof" reformulation stays 
valid in the most general case comprising heavy external momenta with 
arbitrary linear restrictions.

\subsection{Explicitly convergent form for the heavy mass expansion%
}
\label{ss4.7}

As was noted in \cite{partI}, our method of 
expansion leads to expressions containing spurious divergences in 
different terms which cancel in the final result. Let us 
reorganize our asymptotic expansions in such a way as to reexpress 
them in an explicitly convergent form. In \eq{(4.18)}, $L_{\rm eff,
bare}(\varphi )$ is a Lagrangian functional (in terms of 
subsect.~\ref{ss2.1}) of the field $\varphi $ whose coefficients 
(``coupling constants") are functions of the heavy masses. Irrespective 
of the specific values of these functions, the diagrams on the r.h.s.\ 
of \eq{(4.17)} may contain UV divergences. Using \eq{(2.11)}, one can 
explicitly extract the \rop\ on the r.h.s.\ of \eq{(4.17)} and rewrite 
it as:
\begin{equation}\label{(4.19)}
     \bold{As} \. <{\rm T} \exp [L(\varphi ,\Phi ) 
                                + \varphi J]>_{0} 
     = \bold{R} \. <{\rm T} \exp [L_{\rm eff}(\varphi ) 
                                 + \varphi J]>_{0},
\end{equation}
where 
\begin{equation}\label{(4.20)}
     L_{\rm eff}(\varphi ) = \xi ^{-1}[ L_{\rm eff,bare}(\varphi ) ].
\end{equation}

Since the mapping $\xi ^{-1}$ transforms a Lagrangian functional again 
into a Lagrangian functional, $L_{\rm eff}(\varphi )$ can be 
represented as
\begin{equation}\label{(4.21)}
     L_{\rm eff}(\varphi ) 
     = \sum_n \,\int\, dx \, g_{{\rm eff},n}(g,M,\mu ,\epsilon ) 
       j_{n}(x),
\end{equation}
where $j_{n}(x)$ are defined in subsect.~\ref{ss2.1} and $g$ denotes
the set of coupling constants of the original interaction Lagrangian
$L;$ the ``effective couplings" $g_{{\rm eff},n}$
have the form:
\begin{equation}\label{(4.22)}
     g_{{\rm eff},n}(g,M,\mu ,\epsilon ) 
     = {1\over M^{d_n}} \sum_{k=0}^{\infty} 
       g^k\, \sum_{l\ge 0} 
       \left( {\mu\over M} \right)^{l(2\epsilon)} 
       g_{{\rm eff},n,l}(\epsilon),
\end{equation}
where the summation over powers of $g$ reminds one that we are working 
within perturbative framework; $g_{{\rm eff},n,l}(\epsilon )$ have a 
pole singularity at $\epsilon \sim 0$ of a finite order depending on 
$n$ and $l,$ while the integer exponents $d_{n}$ (some which for a few lowest values of $n$ may 
be negative) are determined on dimensional grounds. 

Indeed, the coefficients $g_{{\rm eff},n}$ are polynomial expressions 
built of UV counterterms (each of which is just a polynomial in 
$1/\epsilon $) and heavy knots which are Taylor-expanded in powers of 
their external momenta and $m.$ The resulting loop integrals depend 
only on $M,$ and on dimensional grounds this dependence has the form 
$M^{d'-2\epsilon l'}\times c(\epsilon )$ where $d'$ is an integer and 
$l'$ the corresponding loop number, while $c(\epsilon )$ has a pole 
singularity $1/\epsilon^{l'}$ (the latter information is strictly 
speaking not necessary). The powers of momenta and the light masses 
$m$ resulting from the above Taylor expansion are included into $j$ 
(recall the definition in subsect.~\ref{ss2.1}). Recalling that the UV 
renormalization parameter $\mu$ enters all expressions via integer 
powers of $\mu^{2\epsilon},$ one arrives at \eq{(4.21)} and 
\eq{(4.22)}. Note that the range of summation over $l$ is finite for 
each $n$ and $k.$

Now one can easily convince oneself that the parameters of the 
effective Lagrangian are finite at $\epsilon =0.$ Indeed, the initial 
expression (the l.h.s.\ of \eq{(4.19)}) is finite at $\epsilon=0$ by 
construction in each order of the perturbation theory (in the absence 
of UV divergences in the initial model---recall the assumption at the 
beginning of this section). When transformed into the form of the 
r.h.s.\ of \eq{(4.19)}, it is, of course, still finite but has now the 
form of UV-renormalized (and therefore, finite) loop integrals times 
products of $g_{{\rm eff},n}(g,M,\mu ,\epsilon ),$ in each order of 
the perturbation theory. Taking into account linear independence of 
$j_{n},$ one concludes that all $g_{{\rm eff},n}(g,M,\mu ,\epsilon )$ 
are finite at $\epsilon =0.$

\subsection{\asop\ on UV-divergent diagrams}
\label{ss4.8}

\Eq{(4.2)} was used to define the \asop\ on UV-convergent diagrams. 
However, that definition can be formally extended to the UV-divergent 
case (then, of course, the UV divergences should be regularized; but 
the dimensional regularization which is used at the intermediate steps 
of the \asop\ for products as derived in \cite{partI}, regularizes UV 
divergences as well; therefore, no new regulators are really needed). 
The combinatorial results \eqs{(4.17)}-{(4.20)} remain valid 
though now the effective couplings $g_{{\rm eff},n}(g,M,\mu ,\epsilon 
)$ are not finite at $\epsilon \sim 0$ because they inherit the UV 
divergences of the unrenormalized Green functions on the l.h.s.\ of 
\eq{(4.17)}.

Let $\bold{R} \. \Gamma (k,m,M)$ be a renormalized graph (the 
dependence on the renormalization parameter $\mu $ is implicit; note
that we are dealing here with {\em integrated} graphs, so that
there is no dependence on the loop momenta). The 
\rop\ was defined by \eq{(1.1)}, which can be represented as follows:
\begin{equation}\label{(4.23)}
     \bold{R} \. \Gamma (k,m,M) 
     = \sum_i {\cal Z}_{i}({1/\epsilon},m,M) \Gamma _{i}(k,m,M),
\end{equation}
where $i$ enumerates all the individual terms in the sum in 
\eq{(1.1)}, ${\cal Z}_i$ are products of divergent coefficients of the 
corresponding UV counterterms and are polynomials of all their 
arguments, $\Gamma _{i}$ are the (unrenormalized) diagrams obtained 
from $\Gamma $ by shrinking the corresponding UV-subgraphs to points.

Naively, to expand \eq{(4.23)} one would expand each term on the 
r.h.s.: the factors ${\cal Z}_i$ are polynomials in $m$ and are already 
``expanded", and $\Gamma _{i}$ can be expanded by a straightforward 
application of the definition \eq{(4.2)}. Indeed, all the intricacies 
of the \asop\ are essentially aimed at extracting the terms that are 
non-analytical in light parameters, and the very fact of the 
polynomial dependence of UV counterterms on masses and momenta makes 
one expect that UV renormalization in the MS-scheme and the {\it 
As}-expansion are, in a sense, ``orthogonal" (cf. the analysis of \cite{IV}).

Motivated by the above, we define the \asop\ on renormalized graphs by 
applying the \asop\ \eq{(4.2)} termwise to $\Gamma _{i}$ on the r.h.s.\ 
of \eq{(4.23)} as follows:
\begin{equation}\label{(4.24)}
     \bold{As} \. \bold{R} \. \Gamma (k,m,M) 
     \mathop=^{\rm def} \sum_i {\cal Z}_i({1/\epsilon},m,M) 
     \bold{As} \. \Gamma _{i}(k,m,M).
\end{equation}
We assert that the expansion thus obtained is a correct asymptotic 
expansion for $\bold{R} \. \Gamma (k,m,M).$ This is not obvious, but 
for that matter neither is obvious the UV finiteness of the r.h.s.\ of 
\eq{(4.23)} or even \eq{(1.1)}. 

For a practically oriented reader it may be sufficient to learn 
the following. On the one hand, the final expressions for the 
coefficient functions of short-distance expansions (see 
subsect.~\ref{ss5.3}) imply highly non-trivial cancellations of UV and 
IR divergences between different terms which cannot be understood 
without the underlying theory. Such cancellations were checked by 
explicit three-loop calculations in \cite{alg:83} and in several other 
calculations (see e.g.\ \cite{ope3loops}). On the other hand, from the 
point of view of our methods there is no essential difference between 
the short-distance operator-product expansion and the most general 
Euclidean regimes with heavy masses, and all the non-trivial patterns 
of interaction of UV and IR divergences manifest themselves at the 
three-loop level. One should also take into account that we are 
dealing with relatively primitive although increasingly cumbersome 
integrals of rational functions 
where all the effects are governed by the power counting 
(this point is emphasized and exhibited in the regularization independent 
formalism developed in the parallel series of papers 
\cite{fvt-vvv}--{IV}),
so that there is no room for pathologies. Therefore, the explicit 
calculations of \cite{alg:83} and \cite{ope3loops} provide a sufficient 
evidence in favour of the above assumption.

For a formally oriented reader we note
that as was proved in \cite{gms}, the \rop\ can be represented in such
a form that its UV finiteness becomes obvious. Moreover, 
the reasoning of \cite{IV} demonstrates that the proof of correctness
of the above definition reduces to 
verifying commutativity of two \asop s---one corresponding to the
expansion under consideration, the other being implicit in the 
construction
of UV subtractions along the lines of \cite{gms}. Such commutativity
is essentially due to commutativity of the corresponding formal 
expansions
of the individual factors, and verification of the fact that 
it is indeed
inherited by \asop s on the entire integrands is
a rather straightforward technical excercise. An informal 
discussion of the proofs of \cite{IV} is \cite{III}.

In what follows, we 
will concentrate on the combinatorial structure of the definition 
\eq{(4.24)}. 

(Note that the aggregate operation $\bold{As} \. \bold{R}$ defined in 
\eq{(4.24)} was denoted as $\bold{As}$ and studied as a whole in our 
original publication \cite{IIold'}, and the starting point there was 
the so-called EA-expansion. Such a 
way of reasoning resulted in superfluous complications which are 
avoided in the present exposition.)

\subsection{\asop\ in models with UV divergences}
\label{ss4.9}

If the model described by the interaction Lagrangian $L(\varphi ,\Phi 
)$ possesses UV divergences, then the UV-renormalized generating 
functional of Green functions can be represented as (cf. \eq{(1.34)}):
\begin{equation}\label{(4.25)}
     \bold{R} \. <{\rm T} \exp[L(\varphi ,\Phi ) 
                 + \varphi J]>_{0} 
     = <{\rm T} \exp[L_{R}(\varphi ,\Phi ) + \varphi J]>_{0}.
\end{equation}
with $L_{R}(\varphi ,\Phi )$ defined analogously to \eq{(1.34)}.

Since the \asop\ is defined to act formally on each term of the 
renormalized diagrams, one can apply it to both sides of \eq{(4.25)}. 
Then one can simply use the results obtained for the case of the 
models without UV divergences (see \eqs{(4.17)}-{(4.20)}). Our 
final formulae are as follows:
$$
\bold{R} \. <{\rm T} \exp[L(\varphi ,\Phi ) 
                 + \varphi J]>_{0} 
     \asy{M}{\infty} \bold{As} \. \bold{R} \. 
     <{\rm T} \exp[L(\varphi ,\Phi ) + \varphi J]>_{0}
$$
\begin{equation}\label{(4.26)}
\equiv  \bold{R} \. {\bold{R}^{-1} \. 
     \bold{As} \. \bold{R}} \. <{\rm T} \exp[L(\varphi ,\Phi ) 
                               + \varphi J]>_{0}
     \equiv \bold{R} \. <{\rm T} \exp [L_{\rm eff}(\varphi ) 
                        + \varphi J]>_{0},
\end{equation}
where
\begin{equation}\label{(4.27)}
     \,\quad\quad L_{\rm eff}(\varphi ) 
     = \Delta ^{-1}_{\rm UV} \. 
       \left[ 
          {\rm T}\,  e^{L_{\rm eff,bare}(\varphi)} - 1 
       \right],
\end{equation}
\begin{equation}\label{(4.28)}
     \;\;\;L_{\rm eff,bare}(\varphi ) 
     = \Delta _{as} \. 
       \left[ 
          {\rm T}\,  e^{L_{\rm R}(\varphi,\Phi)} - 1
       \right]_{\Phi=0} ,
\end{equation}
\begin{equation}\label{(4.29)}
     L_{R}(\varphi ,\Phi ) = \Delta _{\rm UV} \. 
     \left[ {\rm T} e^{L(\varphi,\Phi)} - 1 \right].\;
\end{equation}
The algorithms of the three $\Delta $-operations have been described 
in detail in sects.~1.1, 2.8, and 4.4.

\Eq{(4.26)} means that the effects of virtual presence of heavy 
particles on the effective low energy theory of light particles can be 
described by an effective Lagrangian $L_{\rm eff}$ (explicitly given 
by \eqs{(4.27)}-{(4.29)}) to all orders in $1/M.$ Our 
derivation of this result is valid within the MS-like renormalization 
schemes and in all models including those with massless particles like 
QCD.

Note that the \rop s in the original model and in the effective model 
need not be the same; this can be equivalently described by saying 
that the renormalization parameters in the two models may be 
different. In such a case the parameters of $L_{\rm eff}$ will depend 
on {\em both} renormalization parameters.

It should also be recalled that the \asop\ derived in \cite{partI} is 
essentially unique because it leads to series in powers and logarithms 
of the expansion parameters. This uniqueness property (up to the 
choice of renormalization schemes) is inherited by 
\eqs{(4.26)}-{(4.29)}, whatever the guise in which such 
expansions might appear when derived by alternative methods. Our form 
of presentation is dictated by computational convenience.

The property of uniqueness also facilitates the study of 
gauge properties of $L_{\rm eff}$ (cf. \cite{piv-gauge}).

\subsection{Example}
\label{ss4.10}

To get a feeling of how the above formulae work, consider QED with a 
light electron $\psi $ and a heavy muon $\Psi.$ The well-known 
interaction Lagrangian is denoted as $L   (\psi ,\Psi,A)$ where $A$ is 
the photon field. Restricting ourselves to the two-photon sector in 
the 1-loop approximation, we get (fat lines correspond to the muon):
$$
     L_{R}(\psi ,\Psi ,A) 
     = \Delta _{\rm UV} \. 
       \left[ 
          {\rm T}\, e^{L_{\rm QED}(\psi,\Psi,A)} - 1
       \right]
$$
$$
     = \Delta _{\rm UV} \. \Biggl[ \kern2.7cm  \Biggr ]
     + \Delta _{\rm UV} \. \Biggl[ \kern2.7cm  \Biggr ]
     + \ldots 
$$
\begin{equation}
     = \delta Z\, i\,\int\, dx \, 
       \left[
         - {1\over4} F^2_{\mu\nu}(x)
       \right] 
       + \ldots ,
\end{equation}
$$
     L_{\rm bare,eff}(\psi ,A) 
     = \Delta _{\rm as} \. 
       \left[ {\rm T}e^{L_{\rm R}(\psi,A)} - 1 \right]_{\psi=0}
$$
\begin{equation}\label{(4.31)}
     = L_{R}(\psi ,A) 
     + \Delta _{\rm as} \. \Biggl[ \kern2.7cm  \Biggr ] 
     + \ldots ,
\end{equation}
and, finally, 
$$
     L_{\rm eff}(\psi ,A) 
     = \Delta ^{-1}_{\rm UV} \. 
       \left[ {\rm T}\, e^{L_{\rm eff,bare}(\psi ,A)} - 1\right]
$$
\begin{equation}\label{(4.32)}
     = L_{\rm eff,bare}(\psi ,A) 
     + \Delta ^{-1}_{\rm UV} \. 
       \Biggl[ \kern2.7cm  \Biggr ]     + \ldots .
\end{equation}
Since
\begin{equation}\label{(4.33)}
     \Delta ^{-1}_{\rm UV} \. \Biggl[ \kern2.7cm  \Biggr ]
     = - \Delta _{\rm UV} \. \Biggl[ \kern2.7cm  \Biggr ] ,
\end{equation}
we obtain:
$$
     L_{\rm eff}(\psi ,A) = \Delta _{\rm as} \. 
       \Biggl[ \kern2.7cm  \Biggr ]                  
     + \Delta _{\rm UV} \. 
       \Biggl[ \kern2.7cm  \Biggr ]                  
     + \ldots 
$$
\begin{equation}\label{(4.34)}
     = \delta z(M^{2},\mu ^{2}) i\,\int\, dx \, 
       \left[ - {1\over4} F^2_{\mu\nu}(x) \right] + \ldots ,
\end{equation}
where the dots denote other field structures, and
\begin{equation}\label{(4.35)}
     \delta z(M^{2},\mu ^{2}) 
     = {e^2\over 16\pi^2} 
       \left[ c_{1} + c_{2}\log(M^{2}/\mu ^{2}) \right] 
       + O(1/M^{2}).
\end{equation}
Note that cancellation of the UV divergences in the final result is 
obvious even without explicit calculations.

\section{Generalized operator-product expansions}
\label{s5}

In this section we turn to the case when the set of heavy parameters 
includes external momenta. Recall that we regard the diagrams and 
Green functions to be expanded as distributions with respect to the 
heavy external momenta \cite{partI}.

We start by considering in subsects.~\ref{ss5.1}--\ref{ss5.3} 
the case corresponding to 
the usual short-distance operator-product expansion, i.e.\ when there 
are no heavy masses and no linear restrictions on the heavy external 
momenta except the overall momentum conservation. In 
subsect.~\ref{ss5.1} notations are introduced and the \asop\ for 
integrals is derived for the case under study, in subsect.~\ref{ss5.2} 
the global structure of the \asop\ as applied to Green functions is 
studied, and in subsect.~\ref{ss5.3} the final results are presented. 
Modifications due to heavy masses are studied in subsect.~\ref{ss5.4}.

In subsect.~\ref{ss5.5} we impose the so-called ``natural" linear 
restrictions on the heavy momenta, and in subsect.~\ref{ss5.6} the 
corresponding version of the \asop\ and the expansions for Green 
functions are presented. In subsect.~\ref{ss5.7} the structure of the 
general formulae is explained with a simple example.

In subsect.~\ref{ss5.8} we briefly discuss the contact terms in the 
obtained expansions. The section is concluded in subsect.~\ref{ss5.9} 
with a discussion of the most general linear restrictions imposed on 
heavy momenta; it is pointed out that the contact terms in such a case 
may contain the so-called paralocal operators.

\subsection{\asop\ and heavy momenta}
\label{ss5.1}

Let us assume that all the particles of the model are light, but some 
of the external momenta of the Green function to be expanded are 
heavy. 

More precisely, let $j$ numerate the heavy external lines of the 
diagram $\Gamma ,$ then $Q_{j}$ are the corresponding ingoing external 
momenta. There are also some light external momenta collectively 
denoted as $k,$ which are of order $O(\kappa)$ together with the 
masses $m$ (recall that $\kappa $ is our standard notation of the 
expansion parameter \cite{partI}). The fact that $Q$ are heavy is 
formally expressed as
\begin{equation}\label{(5.1)}
     Q=O(1) \quad{\rm as}\quad \kappa  \to 0.
\end{equation}
One cannot assume all $Q_{j}$ to be independent of $\kappa $ because 
of the momentum conservation which reads:
\begin{equation}\label{(5.2)}
     \sum  Q = - \sum  k = O(\kappa ).
\end{equation}
Therefore, let us introduce the ``heavy" components of $Q$ that are 
independent of $\kappa :$
\begin{equation}\label{(5.3)}
     \bar{Q}_{j} \equiv  Q_{j}\vert_{\kappa=0}.
\end{equation}
$Q$ can be represented as:
\begin{equation}\label{(5.4)}
     Q_{j} = \bar{Q}_{j} + q_{j},
\end{equation}
where $q_{j}=O(\kappa )$ are linear combinations of $k.$

The momentum conservation should hold separately for heavy and light 
components, so that:
\begin{equation}\label{(5.5)}
     \sum_j  \bar{Q}_{j} = 0.
\end{equation}
We have assumed that there are no other restrictions on $\bar{Q}$ 
except \eq{(5.5)}. Considering the diagram $\Gamma $ as a distribution 
with respect to the heavy momenta means that we expand expressions of 
the form
\begin{equation}\label{(5.6)}
     \Gamma (k,m,F) 
     \equiv  \,\int\, d\bar{Q}' \,  F(\bar{Q}) \, 
               \int\, dp \, \Gamma (p,\bar{Q}+q,k,m),
\end{equation}
where $F$ is a smooth test function independent of $\kappa ,$ and the 
integration runs over the manifold described by \eq{(5.5)}.

It is convenient to represent $F$ as a vertex attached to the vertices 
corresponding to $Q$ by the lines that are heavy by definition---see 
Fig.~3.

To expand $\Gamma (k,m,F),$ one repeats the reasoning of 
subsects.~4.1--4.4 and arrives at the following equation instead of 
\eq{(4.12)}:
\begin{equation}\label{(5.7)}
     \bold{As} \. \Gamma  
     = \sum_h (\Delta _{\rm as} \. h)\times [\Gamma / h].
\end{equation}
Now there is only one heavy knot shrunk to the point in each term in 
the sum. Its description---owing to the above agreement that the lines 
connecting the $F$-vertex with $\Gamma $ are heavy---coincides with 
that given in subsect.~\ref{ss4.2}: it must be 1PI with respect to 
light lines. But it may be easier to follow the ``fool-proof" recipe 
given at the end of subsect.~\ref{ss4.6}, which remains valid here.

A reasoning similar to that of subsect.~\ref{ss4.5} shows that the 
above formula is also valid in the case of a non-1PI graph $\Gamma ;$ 
in particular, it is valid for disconnected graphs.

\subsection{\asop\ on Green functions}
\label{ss5.2}

Let us now turn to Green functions. Let $H_{j}(x)$ be local products 
of light fields $\varphi (x)$ and their derivatives. Consider the 
following generating functional of Green functions:
\begin{equation}\label{(5.8)}
     <{\rm T}\Bigl\{ \prod_j \tilde{H}_{j}(Q_{j}) e^{{\cal L}} 
             \Bigr\}>_{0},
\end{equation}
where the tilde marks Fourier transforms and ${\cal L}$ is a local 
functional (see the definition in subsect.~\ref{ss2.1}). To obtain 
specific correlators from \eq{(5.8)}, it is sufficient to perform 
suitable variations with respect to the coefficient functions of 
${\cal L}$ and replace ${\cal L}$ by the Lagrangian $L.$ The momenta 
corresponding  to any additional operator insertions are light, i.e.\ 
$O(\kappa ),$ by definition.

Without loss of generality, the test function is introduced as 
follows:
\begin{equation}\label{(5.9)}
     G(F,{\cal L}) \equiv  <{\rm T}\left\{ F\ast {\cal H} e^{{\cal L}} 
                            \right\}>_{0},
\end{equation}
where 
\begin{equation}\label{(5.10)}
     F\ast {\cal H} = \int\, d\bar{Q}' \, 
               F(\bar{Q}) 
               \prod_j \tilde{H}_{j}(\bar{Q}_{j}+q_{j}),
\end{equation}
where the momenta $\bar{Q}_{j}$ and $q_{j}$ are the same as defined in 
subsect.~\ref{ss5.1}.

The sum of all $q_{j}$ should in general not be taken to be zero in 
\eq{(5.9)} in order to get rid of the disconnected diagrams 
contributing to \eq{(5.9)} which normally are of no interest in 
phenomenological applications. If, however, the kinematics of the 
problem require that $\sum q_{j} = 0$ for the connected component, one 
can take the corresponding limit termwise in the final expansion. 
Taking such limits commutes with the expansion procedure unless there 
are connected diagrams like the one shown in Fig.~4 where the wavy 
lines correspond to massless particles and may give rise to an 
infrared divergence. Whether or not the problem under study allows 
such terms can be checked by a straightforward inspection. However, it 
is still possible to regularize such contributions e.g.\ by introducing 
a non-zero mass (which, of course, should be considered as light in 
the expansion procedure); the $\bar{Q}$-dependent part of the 
final expansion as obtained by our methods will be insensitive to the 
light mass structure of the model (see below).

Now, using \eq{(5.7)} and reasoning as in subsect.~\ref{ss4.6}, we obtain the 
following analogue of \eq{(4.17)}:
\begin{equation}\label{(5.11)}
     \bold{As} \. G(F,{\cal L}) 
     = <{\rm T}\left\{ \Delta _{\rm as} \. 
                  \left[ {\rm T} F\ast {\cal H} 
                         e^{{\cal L}}
                  \right] e^{{\cal L}}
               \right\} >_{0}.
\end{equation}
The algorithm for evaluating $\Delta _{\rm as}[{\rm T} F\ast {\cal H} 
\exp {\cal L}]$ is exactly the same as described in 
subsect.~\ref{ss4.6} (recall that according to the graphical 
conventions introduced after \eq{(5.6)}, the lines connecting the 
$F$-vertex with the proper Feynman diagram are heavy by definition; 
note also that the $\delta$-function which expresses momentum 
conservation in the relevant diagrams contains a sum of only light 
momenta due to \eq{(5.5)}, so that a formal expansion in light 
parameters will not affect it; if one wishes, one could introduce a 
formal integration in some of the $q_{j}$ to get rid of it 
completely). One sees that $\Delta _{\rm as}[{\rm T} F\ast {\cal H} 
\exp {\cal L}]$ is a local functional in the sense of 
subsect.~\ref{ss2.1}, so that (cf. subsect.~\ref{ss5.8}):
\begin{equation}\label{(5.12)}
     \Delta _{\rm as} \. 
     \left[ {\rm T} F\ast {\cal H} e^{{\cal L}} \right] 
     = \sum_n C^{\rm bare}_{{\cal H},n}(F,{\cal L}) 
              \tilde{j}_{n}(q),
\end{equation}
where $q = \sum q_{j}$ and 
\begin{equation}\label{(5.13)}
     C^{\rm bare}_{{\cal H},n}(F,{\cal L}) 
     = \,\int\, d\bar{Q}' \, F(\bar{Q}) 
         C^{\rm bare}_{{\cal H},n}(\bar{Q},{\cal L}).
\end{equation}
We have introduced the subscript ``bare" to indicate that the $C$'s 
contain divergences, so that a procedure analogous to what was 
described in subsect.~\ref{ss4.7} is needed, in order to transform 
\eq{(5.11)} to an explicitly convergent form.

To accomplish this, one proceeds as follows:
$$
     \bold{As} \. G(F,{\cal L}) 
     = \sum_n C^{\rm bare}_{{\cal H},n}(F,{\cal L}) 
         <{\rm T} \left\{ \tilde{j}_{n}(q) e^{{\cal L}} 
                  \right\}>_{0} 
$$
$$
     = \sum_n C^{\rm bare}_{{\cal H},n}(F,{\cal L}) 
              {\delta \over \delta {\cal L}_{n}(q) } 
              <{\rm T}\, e^{{\cal L}}>_{0}
$$
$$
     = \sum_n C^{\rm bare}_{{\cal H},n}(F,{\cal L}) 
              { \delta \over \delta {\cal L}_{n}(q) }
              \bold{R} \. 
              < {\rm T} e^{\xi^{-1}[{\cal L}]}
              >_{0}
$$
\begin{equation}\label{(5.14)}
     = \sum_m
       \left\{ 
          \sum_n C^{\rm bare}_{{\cal H},n}(F,{\cal L}) 
                 {\cal Z}_{n,m} 
       \right\} 
       \bold{R} \. 
       <{\rm T} \left\{ \tilde{j}_{m}(q) 
                        e^{\xi^{-1}[{\cal L}]} 
                \right\} >_{0},
\end{equation}
where we have used the fact that $\xi ^{-1}[{\cal L}]$ defined in 
sect.~\ref{s2} is a local functional.

To complete the reasoning, one should take into account that the 
initial expression on which the \asop\ acts may contain UV divergences 
and be renormalized via the \rop, as follows:
\begin{equation}\label{(5.15)}
     G_{\rm R}(F,{\cal L}) 
     \equiv  \bold{R} \. 
        < {\rm T} \left\{ F\ast {\cal H} e^{{\cal L}} \right\} >_{0}.
\end{equation}
The arguments justifying correctness of a straightforward application 
of $\bold{As}$ to renormalized expressions are the same as in 
subsect.~\ref{ss4.8}.

\subsection{Short-distance expansions}
\label{ss5.3}

Now let us exhibit the general structure of the expressions for the 
case when ${\cal L} = L + \varphi J$ where $L$ is a Lagrangian 
functional.

Using the results of sect.~\ref{s3} (cf. \eq{(3.7')}), one represents 
\eq{(5.15)} as:
\begin{equation}\label{(5.16)}
     G_{\rm R}(F, L + \varphi J ) 
    = \sum_a  < {\rm T} 
              \left\{ F\ast {\cal Z}_{{\cal H},a}\ast {\cal J}_{a} 
                      e^{[L_{R}+\varphi J]}
              \right\} >_{0},
\end{equation}
where ${\cal J}_{a}$ are multilocal products of the currents $j_{n}$ 
defined in subsect.~\ref{ss2.1}, ${\cal Z}$ is a ``matrix" of divergent 
coefficients, and the $\ast$'s denote contractions over the heavy 
momenta. $L_{\rm R}$ is defined in \eq{(1.27)}. Applying $\bold{As}$ 
to both sides and using \eq{(5.14)}, one gets:
\begin{equation}\label{(5.17)}
     \bold{As} \. G_{\rm R}(F, L + \varphi J ) 
     = \,\int\, d\bar{Q}' \, 
          F(\bar{Q}) \sum_m c_{m}(\bar{Q},g,\mu ) 
          \bold{R} \. 
          < {\rm T} 
              \left\{ \tilde{j}_{m}(q) 
                     e^{[L+\varphi J]}
              \right\} 
          >_{0},
\end{equation}
where
\begin{equation}\label{(5.18)}
     \int\, d\bar{Q}' F(\bar{Q}) c_{m}(\bar{Q},g,\mu ) 
     \equiv  
     \sum_a \sum_n 
           C^{\rm bare}_{{\cal J}_{a},n}
             (F\ast {\cal Z}_{{\cal H},a},L_{\rm R}) 
           {\cal Z}_{n,m} 
\end{equation}
are linear functionals of the test function $F$ that are finite in the 
limit $\epsilon \to 0$ (which is proved similarly to the reasoning at 
the end of subsect.~\ref{ss4.7}); $g$ are the coupling constants of the 
Lagrangian $L,$ and $\mu $ is the renormalization parameter. The 
expansion \eq{(5.17)} has the form of a familiar operator product 
expansion at short distances. 

It should be stressed that \eq{(5.18)}---despite its 
cumbersome appearance---describes a fully constructive algorithm for 
getting explicit expressions for the coefficient functions 
$C_{m}(\bar{Q},g, \mu ).$ This algorithm is equivalent to the one 
described in \cite{alg:83}, where an assumption on existence and 
properties of the operator-product expansion in the MS-scheme was 
made. Namely, in \cite{alg:83} it was assumed that the coefficient 
functions of such OPE are analytical in masses. One can easily see 
that the expressions \eqs{(5.17)}-{(5.18)} prove that 
assumption: indeed, the coefficient functions as defined above have 
turned out to be independent of $m,$ but the local operators $j$ have 
been allowed to contain non-negative integer powers of $m;$ if the 
local operators are built of only fields, then the coefficient 
functions would become analytical in masses. Non-trivial examples of 
two- and three-loop calculations of coefficient functions using this 
algorithm together with explicit formulae can be found in 
\cite{alg:83} and \cite{ope3loops}.

Finally, note that if one wishes to consider Green functions of 
composite operators, then suitable variations in ${\cal L}$ should be 
introduced into \eq{(5.14)}, \eq{(5.15)} etc., giving rise to 
additional terms in the final result. Thus, if one considers the 
expansion of a correlator of the form
\begin{equation}\label{(5.19)}
     < {\rm T} \left\{ 
                  \tilde{j}_{n}(Q)
                  \tilde{j}_{m}(-Q+q)
                  \tilde{j}_{l}(k) e^{L}
                \right\}
     >_{0},
\end{equation}
then the final expansion will contain the correlators
\begin{equation}\label{(5.20)}
     < {\rm T} \left\{ \tilde{j}_{a}(q)
                       \tilde{j}_{b}(k) e^{L} 
               \right\}
     >_{0} \quad{\rm and}\quad 
     <{\rm T} \left\{ \tilde{j}_{a}(q+k) e^{L} \right\}>_{0}.
\end{equation}
The latter expression is a typical ``vacuum condensate", i.e.\ a vacuum 
expectation value of a local operator (cf. \cite{svz}). It is not 
equal to zero, even within perturbation theory, because the normal 
ordering of local products of fields is not used in the MS-scheme.

\subsection{Effects of heavy particles on operator expansion}
\label{ss5.4}

If there are heavy masses in the model, then the above results will 
get modified in the following way. The expression \eq{(4.12)} for the 
\asop\ will remain valid provided one includes the heavy knots 
corresponding to the $F$-vertex that were described after \eq{(5.7)}. 
Note that the ``fool-proof" recipe of subsect.~\ref{ss4.2} is still 
valid here. Instead of \eq{(5.11)} one will have:
\begin{equation}\label{(5.21)}
     \bold{As} \. G(F,{\cal L})           
     = < {\rm T}
         \left\{ 
            \Delta_{\rm as} \. 
            \left[ {\rm T} F\ast {\cal H} e^{{\cal L}} \right] 
            \exp \Delta _{\rm as} \. 
            \left[ {\rm T}\, e^{{\cal L}} - 1 \right]
         \right\}
       >_{0}.
\end{equation}
And the final result \eq{(5.17)} will take the form
$$
     \bold{As} \. G_{\rm R}(F, L(\varphi ,\Phi ) + \varphi J ) \kern7cm   
$$
\begin{equation}\label{(5.22)}
     = \,\int\, d\bar{Q}' \, F(\bar{Q}) 
                             \sum_m c_{m}(\bar{Q},M,g,\mu ,q) 
     \times \bold{R} \. < {\rm T}
                    \left\{
                       j_{m}(\sum q) 
                       \exp \left[ L_{\rm eff}(\varphi ) 
                                 + \varphi J
                            \right]
                     \right\}
                 >_{0},
\end{equation}
where $L_{\rm eff}(\varphi )$ is defined in \eq{(4.20)}, while $c_{m}$ 
are analytical in $q.$ Note that the operators $j_{m}$ are here built 
of the light fields only, while the currents entering into the 
multilocal operator ${\cal H}$ in the initial expression are allowed 
to contain heavy fields as well.

\subsection{Natural restrictions on heavy momenta}
\label{ss5.5}

Let us turn to the case when there are linear restrictions imposed on 
the heavy external momenta other than the overall momentum 
conservation. Of immediate phenomenological importance are the 
so-called {\em natural restrictions} \cite{IIold"}, \cite{partI}. In 
terms of position space, one can describe them as follows: all the 
``heavy" operator insertions are arranged into several groups, the 
distances within each group tend to zero while the distances between 
groups stay finite (cf. Fig.5). More precise definitions in terms of 
momentum representation are presented below.

Consider \eq{(5.8)} and let the ``heavy" operator insertions $H_{j}$ 
(which are allowed to be built of both  heavy and light fields) be 
divided into several non-intersecting groups (numerated by $\lambda $) 
in such a way that the heavy momentum conservation holds separately 
within each group:
\begin{equation}\label{(5.23)}
     \sum_{j\in\lambda } Q_j = O(\kappa ),
     \quad\quad {\rm for\ each\ \ } \lambda .
\end{equation}
In terms of the independent momenta $\bar{Q}:$
\begin{equation}\label{(5.24)}
     \sum_{j\in\lambda } \bar{Q}_{j} = 0, 
     \quad\quad {\rm for\ each\ \ } \lambda .
\end{equation}
For each group it is natural to introduce a separate test function 
$F_{\lambda }(\bar{Q}_{\lambda })$ (where $\bar{Q}_{\lambda }$ denotes 
the set of all $\bar{Q}_{j}$ for $j\in \lambda $), and the 
corresponding integration $d\bar{Q}'_{\lambda }$ over the manifold 
described by \eq{(5.24)}. Now eqs.\eq{(5.9)} and \eq{(5.10)} are 
replaced by:
\begin{equation}\label{(5.25)}
     G({F_{\lambda }},{\cal L}) 
     \equiv  < {\rm T} \left\{ 
        \prod_\lambda F_{\lambda }\ast {\cal H}_{\lambda } 
                     e^{{\cal L}(\varphi ,\Phi )}
     \right\} >_{0},
\end{equation}
where 
\begin{equation}\label{(5.26)}
     F_{\lambda }\ast {\cal H}_{\lambda } 
     = \int\, d\bar{Q}'_{\lambda }F_{\lambda }({\bar Q}_{\lambda })
       \prod_{j\in\lambda } \tilde{H}_{j}(\bar{Q}_{j}+q_{j}).
\end{equation}
To use graphical representation, one introduces an $F$-vertex of the 
same type as described after \eq{(5.6)} for each of the test functions 
$F_{\lambda }.$

\subsection{Generalized operator expansions}
\label{ss5.6}  

The \asop\ on a diagram $\Gamma $ contributing to \eq{(5.25)} will 
have the form
\begin{equation}\label{(5.27)}
     \bold{As} \. \Gamma 
     = \sum_{\{h_j\}}
         \left(
           \prod_i \Delta _{\rm as} \. h_{i}
         \right) \times 
         \left[ 
            \Gamma  / \prod_i h_{i}
         \right]_{M=\infty,F=0}
\end{equation}
(cf. \eq{(4.12)}), where the heavy knots $h$ are 1PI with respect to 
light lines and must contain at least one $F$-vertex or heavy line. 
Setting $M=\infty $ and $F=0$ is an expression of the fact that all 
the ``heavy" elements ($F$-vertices and heavy lines) must be contained 
within the $h$'s.

Applying $\bold{As}$ to \eq{(5.25)}, one obtains the following 
expression instead of \eq{(5.21)}:
$$
     \bold{As} \. G({F_{\lambda }}, {\cal L}(\varphi ,\Phi )){}
$$
\begin{equation}\label{(5.28)}
     = \sum_{\{\Lambda\}}
        < {\rm T} \bigl\{
             \prod_{\Lambda\in\{\Lambda\}} \Delta _{\rm as} \. 
             \left[ {\rm T} \prod_{\lambda\in\Lambda} 
                         F_{\lambda }\ast {\cal H}_{\lambda } 
                         e^{{\cal L}(\varphi ,\Phi )}
             \right] 
     \times \exp \Delta _{\rm as} \. 
        \left[ {\rm T} e^{{\cal L}(\varphi ,\Phi )} - 1 \right] 
      \bigr\} >_{0},
\end{equation}
where $\{\Lambda \}$ denotes  a splitting of the set of all $\lambda $ 
into non-intersecting subsets numerated by $\Lambda ,$ so that
\begin{equation}\label{(5.29)}
     \prod_{\Lambda\in\{\Lambda\}} 
     \prod_{\lambda\in\Lambda}  
     F_{\lambda }\ast {\cal H}_{\lambda } 
     = \prod_\lambda F_{\lambda }\ast {\cal H}_{\lambda },
\end{equation}
and the first summation on the r.h.s. of \eq{(5.28)} runs over all such 
splittings $\{\Lambda \}.$

The action of $\Delta _{\rm as}$ can be most easily described using 
the ``fool-proof" recipe presented in subsect.~\ref{ss4.2}: $\Delta 
_{\rm as}$ nullifies unconnected diagrams as well as all those 
connected ones which result in meaningless expressions like 1/0 when 
formally Taylor-expanded in light parameters. The remaining diagrams 
are exactly 1PI with respect to light lines and get formally 
Taylor-expanded in the light parameters. 

Each $\Delta _{\rm as}$ results in a local functional (cf. 
\eq{(5.12)}), so that \eq{(5.28)} can be represented as
$$
     \bold{As} \. G(\{F_{\lambda }\}, {\cal L}(\varphi ,\Phi )) {} 
$$
$$
     = \sum_{\{\Lambda\} }< {\rm T} \bigl\{ 
       \prod_{\Lambda\in\{\Lambda\}}\left[ 
          \sum_{a_\Lambda } 
              C_{\Lambda ,a_{\Lambda },\rm bare}
              ( \{F_{\lambda }\}_{\lambda \in\Lambda },M,{\cal L},
                \{q_{j}\}_{j\in\Lambda },\epsilon ) 
              \tilde{j}_{a_\Lambda} 
              \left(\sum_{j\in\Lambda}q_{j}\right)
       \right]
$$
\begin{equation}\label{(5.30)}
      \times \exp {\cal L}_{\rm eff,bare}(\varphi ) \bigr\} >_{0},
\end{equation}
where each $C$ is a coefficient function which is independent of 
fields and is at the same time a linear functional with respect to 
each $F_{\lambda }.$ The dependence on $q_{j}$ is analytical.

To obtain an expansion for the renormalized Green functions in an 
explicitly convergent form one should---as has been done in all the 
special cases considered above---``sandwich" the \asop\ between 
$\bold{R}$ and $\bold{R}^{-1}$ (cf. \eq{(4.26)}). Further reasoning is 
similar to that of subsects.4.7 and 5.2--3: one should first get rid of 
the \rop\ using the generalized Zimmermann identities of 
subsect.~\ref{ss3.1}, then apply the \asop, and, finally, use the 
identities of subsect.~\ref{ss3.2} to extract the \rop\ and thus 
transform the expansion to an explicitly convergent form. In this way 
one arrives at the following final result:
$$
     \bold{As} \. \bold{R} \. G ( \{F_{\lambda }\}, L(\varphi ,\Phi ) 
     + \varphi J ){}
$$
$$
     = \sum_{\{\Lambda \}} \bold{R} \. < {\rm T} \bigl\{ 
       \prod_{\Lambda \in\{\Lambda \}} \left[ 
          \sum_{a_\Lambda } c_{\Lambda ,a_\Lambda } \left(
             \{F_{\lambda }\}_{\lambda \in \Lambda },M,g,
             \{q_{j}\}_{j\in \Lambda }
          \right) 
          \tilde{j}_{a_\Lambda } \left( 
              \sum_{j\in \Lambda } q_{j}
          \right)
       \right]
$$
\begin{equation}\label{(5.31)}
     \times  \exp \left[
        L_{\rm eff}(\varphi ) + \varphi J
     \right] \bigr\} >_{0},
\end{equation}
which should be compared with \eq{(5.17)}; $L_{\rm eff}$ is given by 
\eq{(4.20)} and has the form of \eq{(4.21)}.

\subsection{Example}
\label{ss5.7}

To clarify the above general expansion, consider the case of two 
groups of two currents each (the corresponding phenomenological 
problem is the deep inelastic scattering of two deeply virtual photons 
\cite{DISlight}). There are two heavy momenta $\bar{Q}_{1}$ and 
$\bar{Q}_{2},$ and two test functions, $F_{\rm AB}(\bar{Q}_{1})$ and 
$F_{\rm CD}(\bar{Q}_{2}).$ The expression to be expanded is:
$$
     \int\, \,\int\, d\bar{Q}_{1}\, d\bar{Q}_{2} 
       F_{\rm AB} (\bar{Q}_{1}) \, F_{\rm CD}(\bar{Q}_{2}) 
$$
\begin{equation}\label{(5.32)}
     \times  \bold{R} \. < {\rm T} \left\{
         \tilde{j}_{\rm A} (\bar{Q}_{1})
         \tilde{j}_{\rm B} (-\bar{Q}_{1}+q_{1})  
         \tilde{j}_{\rm C} (\bar{Q}_{2})
         \tilde{j}_{\rm D} (-\bar{Q}_{2}+q_{2})
         e^{[L(\varphi ,\Phi )+\varphi J]}
     \right\} >_{0},
\end{equation}
where $q$ are introduced to get rid of the non-connected components. 
Its expansion is as follows. There are two sets $\{\Lambda \}$ over 
which the summation in \eq{(5.31)} runs: $\{\rm (A,B),(C,D)\}$ and 
$\{\rm A,B,C,D\}.$ The first of them corresponds to that term in the 
expansion where the two pairs of currents are ``shrunk" into (linear 
combinations of) local operators separately; the second corresponds to 
a single sum of local operators. The resulting expansion reads (cf. 
Fig.6):
$$
     \int \,\int\, d\bar{Q}_{1} \, d\bar{Q}_{2} \, 
      F_{\rm AB}(\bar{Q}_{1}) F_{\rm CD}(\bar{Q}_{2}) 
$$
$$
     \times  \biggl( \sum_a \sum_b c_{{\rm AB},a} ( \bar{Q}_{1},M,g,\mu ) 
                            c_{{\rm CD},b} ( \bar{Q}_{2},M,g,\mu ) 
$$
$$
     \times  \bold{R} \. < {\rm T} \left\{
          \tilde{j}_{a}(q_{1}) 
          \tilde{j}_{b}(q_{2}) 
          \exp[L_{\rm eff}(\varphi ) + \varphi J]
     \right\} >_{0}
$$
\begin{equation}\label{(5.33)}
     + \sum_a c_{{\rm ABCD},a} ( \bar{Q}_{1},\bar{Q}_{2},M,g,\mu ) 
       \bold{R} \. < {\rm T} \left\{ 
           \tilde{j}_{a}(q_{1}+q_{2})
           \exp[L_{\rm eff}(\varphi )+\varphi J]
       \right\} >_{0}
    \biggr).
\end{equation}

\subsection{Contact terms in operator expansions}
\label{ss5.8}

The expansions that we have derived are valid for arbitrary test 
functions $F(\bar{Q})$. For example, if there are two independent 
heavy momenta, $\bar{Q}_{1}$ and $\bar{Q}_{2},$ then the expansion 
will in general contain terms proportional to $\delta (\bar{Q}_{1}),$ 
$\delta (\bar{Q}_{2}),$ $\delta (\bar{Q}_{1}+\bar{Q}_{2}),$ $\delta 
(\bar{Q}_{1})\delta (\bar{Q}_{2})$ etc. (for a discussion of the role 
of such terms see \cite{partI}). Let us explain how such terms are 
generated by the \asop.

Within the context of subsect.~\ref{ss5.2} (no heavy masses etc.), 
consider the case corresponding to the product of two currents in 
\eq{(5.8)}:
\begin{equation}\label{(5.34)}
     F\ast {\cal H} 
     = \int\, d\bar{Q} \, F(\bar{Q}) H(\bar{Q}+q)H(-\bar{Q}).
\end{equation}
The non-trivial terms are generated from the following expression (cf. 
\eq{(5.12)}):
\begin{equation}\label{(5.35)}
     \Delta _{\rm as} \. \left[
         {\rm T} F\ast {\cal H} e^{L}
     \right] 
     = \sum_n C^{\rm bare}_{{\cal H},n}(F,L) 
       \tilde{j}_{n}(q),
\end{equation}
where $L$ is the interaction Lagrangian. 

There are two classes of terms generated by ${\rm T} F\ast {\cal H} 
e^{L}$ that give non-zero result after application of $\Delta _{\rm 
as}.$ The first class contains the terms whose graphs remain connected 
after the $F$-vertex is deleted. Such terms can be represented as:
\begin{equation}\label{(5.36)}
     \mu ^{-2\epsilon }\,\int\, d^D\bar{Q} \, 
     F(\bar{Q}) f(\bar{Q},p_{1},\ldots,p_{n},m,\mu )
     \times  \delta (q + \sum p_{j})
     \prod_j (\tilde{\varphi }(p_{j})\, d^Dp_{j} ) ,
\end{equation}
where the Feynman amplitude $f$ depends on the masses $m$ of the 
theory, contains one factor $\mu^{2\epsilon }$ per each loop, and the 
$\delta$-function that expresses the momentum conservation is shown 
explicitly. The action of $\Delta _{\rm as}$ on \eq{(5.36)} consists 
in Taylor-expanding $f$ in masses and $p_{j}.$ The result has the 
form:
$$
     \Delta _{\rm as} \. \hbox{\eq{(5.36)}} 
     = \int\, d^D\bar{Q} F(\bar{Q}) 
       \sum_a f_{a}(\bar{Q},\mu )\kern4.5cm
$$
\begin{equation}\label{(5.37)}
     \times  \left[ 
         \mu ^{-2\epsilon } \,\int\, {\cal P}_{a}(m,p_1,\ldots,p_n) 
         \delta (q + \sum p_{j}) 
         \prod_j (\tilde{\varphi }(p_{j})d^Dp_{j})
     \right] .
\end{equation}
The square-bracketed term in \eq{(5.37)} corresponds to $\tilde{j}$ in 
\eq{(5.12)}, and the dependence of $f_{a}(\bar{Q},\mu )$ on $\bar{Q}$ 
and $\mu $ is as follows:
\begin{equation}\label{(5.38)}
     f_{a} (\bar{Q},\mu ) 
     = \bar{f}_{a} (\bar{Q}, \biggl({m^2\over Q^2}\biggr)^\epsilon  ).
\end{equation}
In the $l$-loop approximation, $\bar{f}_{a}$ is a polynomial of order 
$l$ in its second parameter, while its dependence on the first 
parameter is determined by power counting and covariance properties. 

The second class contains the heavy knots that consist of two parts 
connected only via the $F$-vertex (cf.\ Fig.~7). If there are no heavy 
masses, then there is only one such subgraph that is not nullified by 
the operation $\tau $---it consists of the two vertices corresponding 
to the ``heavy" operators $H$ (see \eq{(5.34)}) connected to the 
$F$-vertex. The corresponding contribution to \eq{(5.35)} has the 
following form (for definiteness we assume that $H(x)=\varphi 
(x)^{2}$):
$$
     \Delta _{\rm as} \. [ \int\, dQ\,F(Q)\,\delta (q+Q+p_{1}+p_{2})
                                            \delta (-Q+p'_{1}+p'_{2})\kern2cm
$$
$$
     \kern3cm\times  \tilde{\varphi } (p_{1}) 
       \tilde{\varphi } (p_{2})
       \tilde{\varphi }(p'_{1})
       \tilde{\varphi }(p'_{2}) \, dp'_{1} \ldots dp'_{2} ]
$$
$$
     = \tau [ dQ \, F(Q) \, \delta (-Q+p'_1+p'_2) ] \kern5.5cm
$$
\begin{equation}\label{(5.39)}
     \kern1.5cm\times  \delta ( q + p_{1} + p_{2} + p'_{1} + p'_{2} )
     \tilde{\varphi }(p_{1})
     \tilde{\varphi }(p_{2})
     \tilde{\varphi }(p'_{1})
     \tilde{\varphi }(p'_{2}) \, dp_1 \ldots dp'_2 ,
\end{equation}
where the two $\delta$-functions in the first line are inherited from 
the two operators $H$ (cf.\ \eq{(2.1)}), and we have extracted a 
$\delta$-function in the last line that depends only on the light 
momenta. We have deliberately not performed the integration over $Q$ 
on the r.h.s.\ because $\tau $ should Taylor-expand the 
square-bracketed expression in $p'_i$ and this can be done by applying 
$\tau $ directly to the $\delta$-function. The result has the form
\begin{equation}\label{(5.40)}
     \sum_a \,\int\, dQ \, 
     [F(Q) \delta^{(\alpha )} (Q)] \times  j_{\alpha }(q),
\end{equation}
and one sees that the ``bare" coefficient functions in \eq{(5.12)} 
receive $\delta$-functional contributions. Such 
contributions---``dressed" by the appropriate divergent renormalization 
factors (cf.\ \eq{(5.18)})---are finally inherited by the coefficient 
functions of the short-distance expansion \eq{(5.17)}. It may be not 
quite obvious that the coefficient functions thus constructed will be 
integrable around $Q\sim 0.$ However, cancellations between different 
terms can be traced if one performs the expansion procedure explicitly 
starting from the \asop\ for products of singular functions prior to 
performing the integrations of $\delta$-functions described in 
subsect.~\ref{ss4.3}. Alternatively, an interested reader can verify 
validity of our recipes by straightforward calculations in a simplest 
situation, e.g. within the model $\varphi ^{3}$ in two dimensions for 
the currents $H(x)=\varphi ^{2}(x)$ in one loop approximation.

\subsection{General linear restrictions on heavy momenta and paralocal 
operators}
\label{ss5.9}

Consider the following operator product instead of \eq{(5.10)}:
\begin{equation}\label{(5.41)}
     F\ast {\cal H} = \int\, d\bar{Q}' \,  F(\bar{Q})                       
     \left[ 
        \delta (q + \sum_j \bar{Q}_{j}) 
        \prod_\lambda  \delta ( q_{\sigma } 
                              + \sum_j c_{\lambda ,j} 
                                \bar{Q}_{j} 
                              )
     \right]\left[ 
        \prod_j \tilde{H}_{j} (\bar{Q}_{j}+q_{j})
     \right],
\end{equation}
where the $\delta$-functions are introduced to impose linear 
restrictions on the heavy momenta. With a special choice of 
$c_{\lambda ,j}$ one reproduces the restrictions \eq{(5.24)}. Examples 
of more general restrictions are presented in Fig.~8. We are not aware 
of any phenomenological applications where such restrictions might 
emerge naturally. Therefore, we offer only a few comments concerning 
the most general case \eq{(5.41)}.

First, it should be stressed that the expansion procedure as described 
by \eq{(5.11)} remains fully correct and well-defined provided the 
``fool-proof" recipe for $\Delta _{\rm as}$ from subsect.~\ref{ss4.6} is 
used. No simple graphical description for heavy knots exists, though.

Second, the representation \eq{(5.12)} should be replaced by a more 
general one: the sum on the r.h.s. may now contain operators that can 
not be interpreted as local operators and differ from \eq{(2.1)} by 
additional $\delta$-functions besides the one expressing momentum 
conservation. Such operator monomials can be called {\em paralocal 
operators} \cite{IIold"}. A simple example of a paralocal operator in 
position-space representation may be as follows:
\begin{equation}\label{(5.42)}
     {\rm P}(z) = \int \, \prod_\lambda  \, d\xi _{\lambda } 
     \prod_j \varphi (z + \sum_\lambda  \xi _{\lambda }c_{\lambda  j}),
\end{equation}
but more complicated patterns are possible: note in this respect that 
a natural position-space representation for the multilocal operators 
which are a special case of paralocal ones, is
\begin{equation}\label{(5.43)}
     J_{1}(z_{1})..J_{n}(z_{n}),
\end{equation}
where all $J$ are local operators.

The renormalization properties of paralocal operators can be studied 
most easily using the techniques of \cite{gms} (also \cite{III}) 
and are similar to those of (multi-) local ones: a paralocal operator 
renormalizes via paralocal ones.
Zimmermann identities (including the inverted ones) can also be 
generalized to this case. Therefore, expansions in explicitly 
convergent form for the case of general non-natural restrictions on 
heavy momenta can be obtained in a manner completely similar to the 
case of natural restrictions.

\hbox{}

\hbox{}

\hbox{}

\centerline{{\large\bf Conclusions}}

We have developed a simple combinatorial technique for studying 
global exponentiation properties of the \asop\ and 
obtained Euclidean asymptotic expansions for MS-re\-nor\-ma\-lized Green 
functions of arbitrary local operators in arbitrary models. 
The expansions are true infinite asymptotic series 
that run in powers and 
logs of the expansion parameter. The obtained expansions exhibit 
perfect factorization of heavy and light dimensional parameters 
(masses and external momenta) which means e.g.\ that the coefficient 
functions of operator expansions are analytical in light masses and 
momenta. It should be stressed that the expansions are valid in models 
with massless particles like QCD, and are most convenient for 
practical calculations. Uniqueness of such expansions 
(cf.\ \cite{partI}) 
greatly facilitates study of their properties in gauge models.

\hbox{}

\hbox{}

\hbox{}

{}
{}

{}

\noindent{\large\it Acknowledgments}.

The authors thank A.~N.~Kuznetsov 
for a critical reading of 
early versions of the manuscript. One of the authors (F.~T.) is 
indebted to A.~V.~Radyushkin for his continuous stimulating interest in 
this work, to W.~A.~Bardeen and R.~K.~Ellis for the hospitality at 
the Theory division of Fermilab 
where the final version of the manuscript was 
completed, and to J.~C.~Collins 
for a careful reading of the manuscript and
supplying a list of misprints and comments.

\newpage\thispagestyle{myheadings}\markright{}

\newpage\thispagestyle{myheadings}\markright{}

\noindent{\large\bf Figure captions}

{}

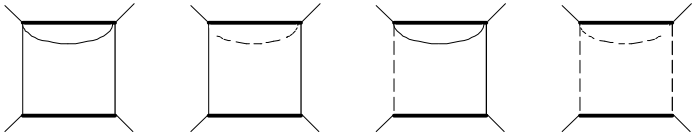
Fig.~1. A diagram (a) and its heavy knots (a)--(d). All the external 
momenta are light. The fat lines correspond to propagators of heavy 
particles. Various IR-subgraphs are shown with dashed lines in 
(b)--(d).

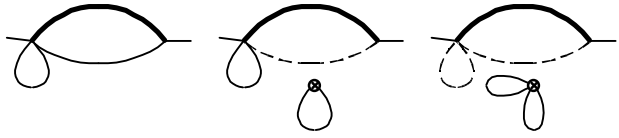
Fig.~2. (a) An example of a diagram with a "light" tadpole subgraph 
(the notations are the same as in Fig.~1). The tadpole corresponds to a 
non-expandable global factor. The \asop\ should not affect such 
factors. (b) and (c) are two IR-subgraphs. The contribution 
corresponding to the IR-subgraph shown in (b) is automatically set to 
zero by the operation $\tau$ due to the properties of dimensional 
regularization, so that the tadpole can be factored out in the result 
produced by the \asop, as expected.

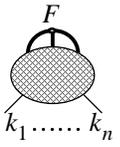
Fig.~3. A graphical representation of the test function corresponding 
to the heavy external momenta. The lines connecting the $F$-vertex 
with the rest of the diagram are heavy by definition.

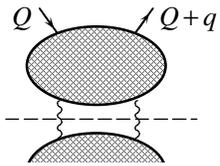
Fig.~4. If the wavy lines correspond to massless scalar particles then 
setting $q=0$ would result in an IR divergence in $D=4$.

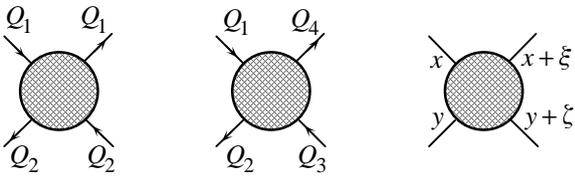
Fig.~5. An example of kinematics of heavy momenta with linear 
restrictions. (a) differs from (b) by the restriction $Q_3=-Q_2$. 
(c) shows the corresponding position space picture: $\xi,\zeta\to 0$.

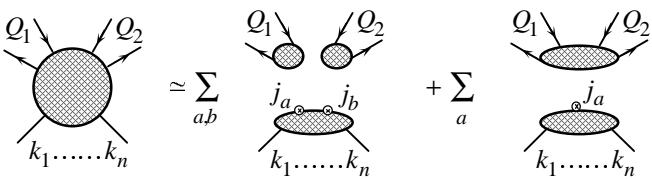
Fig.~6. A diagrammatic illustration of \eq{(5.33)}. The two terms on 
the r.h.s.\ correspond to two ways of cutting off the heavy knots 
corresponding to the heavy external momenta.

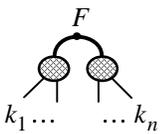
Fig.~7. Graphical representation of the heavy knots contributing to 
contact terms. Each blob corresponds to a subgraph that is 1PI with 
respect to light lines. If there are no heavy masses then each blob is 
just 1PI and is nullified by $\tau$ unless it consists of a single 
vertex.

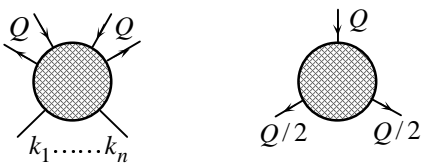
Fig.~8. Examples of ``non-natural" restrictions on heavy momenta. (a) 
differs from Fig.~6 by an additional restriction $Q_1=Q_2$.

\end{document}